\documentstyle[aps,epsfig,preprint]{revtex}

\begin{document}

\newcommand{\pathdefs}{/home/treiber/tex/definitions}
\newcommand{\pathfigs}{.}

\newcommand{\erw}[1]{\mbox{$\langle #1 \rangle$}} 

\tighten
\onecolumn

\title{Congested Traffic States in Empirical Observations
       and Microscopic Simulations}
\author{Martin Treiber, Ansgar Hennecke, and Dirk Helbing}
\address{II. Institute of Theoretical Physics, University of Stuttgart,
         Pfaffenwaldring 57, D-70550 Stuttgart, Germany\\
{\tt http://www.theo2.physik.uni-stuttgart.de/treiber/, helbing/} 
}

\date{\today}
\maketitle

\begin{abstract}
We present data from several German freeways showing different
kinds of congested traffic forming near road inhomogeneities,
specifically lane closings,
intersections, or uphill gradients. 
The states are localized or extended,
homogeneous or oscillating. 
Combined states are observed as well, 
like the coexistence of moving localized clusters
and clusters pinned at road inhomogeneities,
or regions of oscillating congested
traffic upstream of nearly homogeneous congested traffic.
The experimental findings are consistent with a recently proposed
theoretical phase diagram for traffic near on-ramps 
[D. Helbing, A. Hennecke, and M. Treiber, 
Phys. Rev. Lett. {\bf 82}, 4360 (1999)].
We simulate these situations with a novel continuous microscopic
single-lane model, {the ``intelligent driver model'' (IDM), using the
empirical boundary conditions.} All observations,
including the coexistence of states, {are qualitatively
reproduced by describing inhomogeneities with} 
local variations of one model parameter. 
We show that the results of the microscopic model
can be understood by formulating the
theoretical phase diagram for {bottlenecks in a more general way.}
In particular, a local drop of the road capacity induced
by parameter variations has {practically} the same effect
as  an on-ramp.

\end{abstract}



\section{Introduction}

Recently, there is much interest in the 
dynamics of traffic breakdowns 
behind bottlenecks
\cite{Hall1,Daganzo-ramp,Kerner-rehb96-2,Kerner-sync,Nagatani-sync,Persaud,Kerner-wide,sync-Letter,TSG-science,Lee,Lee99,Lee-emp,Daganzo-ST,Phase,Kerner-Transp}.
Measurements of traffic breakdowns on various freeways
in the USA \cite{Daganzo-ST,Hall1,Daganzo-ramp,Persaud},
Germany, \cite{Kerner-rehb96-2,Kerner-sync,Kerner-rehb98,numerics}, 
Holland \cite{Helb-book,Helb-emp97,GKT-scatter,Smulders1,Vladi-98}, 
and Korea \cite{Lee-emp}
suggest that many dynamic aspects are universal and 
therefore accessible to
a physical description.
One common property is the capacity drop (typically of the order of
20\%) associated with a breakdown
\cite{Hall1,Daganzo-ST,Persaud}, 
which leads to hysteresis effects and is the basis of
applications like dynamic traffic control with the aim of
avoiding the breakdown.
In the majority of cases, traffic breaks down upstream of a
bottleneck and the congestion has a stationary donstream
front at the bottleneck. The type of bottleneck, e.g.,
on-ramps
\cite{Daganzo-ramp,Daganzo-ST,Lee-emp,Kerner-sync}, 
lane closings, or uphill gradients \cite{numerics}, seems not to be 
{of importance}.
Several types of congested traffic have been found, among them
extended states with a relatively high traffic flow. These states,
sometimes referred to as ``synchronized traffic'' \cite{Kerner-sync},
can be more or less homogeneously flowing, or show distinct
oscillations in the time series of detector data 
\cite{Kerner-rehb96-2}.
Very often, the congested traffic flow is, apart from fluctuations,
homogeneous near the
bottleneck, but oscillations occur further upstream \cite{Kerner-wide}.
In other cases, one finds isolated stop-and-go waves
that propagate in the upstream direction with a characteristic
velocity of about 15 km/h \cite{Kerner-rehb96,Kerner-rehb98}. 
Finally, there is also {an}  observation of
a traffic breakdown to a pinned localized
cluster near an on-ramp \cite{Lee-emp}.

There are several possibilities to {delineate traffic mathematically,} 
among them macroscopic models describing the dynamics in
terms of aggregate quantities like density or flow
\cite{Lighthill-W,Cremer-93,KK-94,GKT,Lee}, and microsopic
models describing the motion of individual vehicles.
The latter include continuous-in-time models (car-following models)
\cite{Newell,Gipps81,Wiedemann,Bando,Krauss-traff98,Tilch-GFM,MITSIM,Howe,coexist,TGF99-Treiber},
and cellular automata 
\cite{Biham,Cremer-multi,Nagel-S,Barlovic,Helb-sblock,Wolf-Galilei}.
Traffic breakdowns behind bottlenecks have been simulated with the
non-local, gaskinetic-based traffic model
(GKT model) \cite{GKT}, the 
K\"uhne-Kerner-Konh\"auser-Lee model (KKKL model)
\cite{Kuehne,KK-94,Lee}, and with 
a new car-following model, which will
be reported below.

For a direct comparison with empirical data,
one would prefer car-following
models. {As the} position and velocity of each car is known in such
models, one can reconstruct the way how data are obtained by usual 
induction-loop detectors.
To this end, one introduces ``virtual'' detectors recording
passage times and velocities of crossing vehicles and compares this
ouput with the empirical data.
Because traffic density is not a primary variable,
this avoids the problems associated {with} determining the traffic density
by temporal averages \cite{Helb-book}.

The present study 
refers to publication \cite{Phase}, where, based on
a gas-kinetic-based macroscopic simulation model, it has been concluded
that there should be five different congested traffic states on freeways
with inhomogeneities like on-ramps. The kind of congested state
depends essentially on the inflow into the considered freeway section
and on the ``bottleneck strength'' characterizing the inhomogeneity.
This can be summarized by a phase diagram depicting the kind of
traffic state as a function of these two parameters.
A similar phase diagram has been obtained for the KKKL model \cite{Lee99}.
The question is, whether this finding
is true for some macroscopic models only, or universal for a larger class of 
traffic models and confirmed by empirical data.

The relative positions of some of the traffic states in this phase space
has been qualitatively
confirmed for a Korean freeway \cite{Lee-emp}, but only one type of 
extended state has been measured there. Furthermore, this state
did not have  the characteristic properties 
of extended congested traffic on most other freeways,
cf. e.g., \cite{Kerner-sync}.
It is an open question
to confirm the relative position of the other states.
Moreover, to our knowledge, there are no direct simulations of the different
breakdowns using empirical data as boundary conditions,
neither with microscopic nor with macroscopic models.

Careful investigations
with a new follow-the-leader model (the IDM model) 
\cite{coexist,TGF99-Treiber} show that
(i) the conclusions of Ref \cite{Phase} are also valid for certain
  microscopic traffic models (at least deterministic models with
  a metastable density range),
(ii) the results can be systematically transferred to more general
  (in particular flow-conserving) kinds of bottlenecks, and a formula
  allowing to quantify the bottleneck strength is given [see Eq. (15)], 
(iii) the existence of all predicted traffic states is empirically
  supported,
and finally,
(iv) all different kinds of breakdowns can be simulated with the IDM
  model with {\it empirically measured} boundary conditions,
  varying only one parameter (the average time headway T), which
  is used to specify the capacity of the stretch.

The applied IDM model 
belongs to the class of deterministic
follow-the-leader models like the optimal velocity model by Bando et al.
\cite{Bando}, but it has the following advantages:
(i) it behaves accident-free because of the dependence on the
  relative velocity,
(ii) for similar reasons and because of metastability, it shows
  the self-organized characteristic traffic constants demanded 
  by Kerner et al. \cite{KK-94} (see Fig. 4), hysteresis effects
\cite{Treiterer,Daganzo-ST}, and complex states
\cite{Kerner-rehb96-2,Kerner-wide},
(iii) all model parameters have a reasonable interpretation, are known to 
  be relevant, are empirically measurable, and have the expected
  order of magnitude \cite{TGF99-Treiber},
(iv) the fundamental diagram and the stability properties of the model
  can be easily (and separately) calibrated to empirical data,
(v) it allows for a fast numerical simulation,
and
(vi) an equivalent macroscopic version of the model is known
\cite{TGF99-Hennecke}, which is
  not the case for most other microscopic traffic models.

These aspects are discussed in Sec.~II, while Sec.~III is not
model-specific at all. Section III presents ways to specify and quantify
bottlenecks, as well as the traffic states resulting for different traffic
volumes and bottleneck strengths. The analytical expressions for the phase
boundaries of the related phase diagram allow to conclude that similar
results will be found for any other traffic model with a stable,
metastable, and unstable density range. Even such subtle features like
tristability first found in macroscopic models 
\cite{Lee99,TGF99-Treiber} are observed. It would be certainly interesting
to investigate in the future, whether the same phenomena are also found
for CA models or stochastic traffic models like the one by Krauss
\cite{Krauss-traff98}.

Section~IV discusses empirical data  using
representative examples out of a sample
of about 100 investigated breakdowns. 
Thanks to a new method for presenting
the cross section data (based on a smoothing and interpolation procedure),
it is possible to present 3d plots of the empirical density or average
velocity as a function of time {\it and} space. This allows a good imagination
of the traffic patterns and a direct visual comparison with simulation
results.
In the IDM microsimulations, we used a very
restricted data set, namely only the measured flows 
and velocities at the upstream and downstream boundaries omitting the
data of the up to eight detectors in between.
Although the simulated sections were up to 13
km long and the boundaries were typically outside of congestions, 
the simulations reproduced 
qualitatively  the sometimes very complex
observed collective dynamics.

All in all the study supports the idea of the suggested phase diagram of
congested traffic states quite well and suggests ways to simulate real
traffic breakdowns at bottlenecks with empirical boundary conditions.

\section{The Microscopic ``Intelligent Driver Model'' (IDM)}
For about fifty years now, researchers model
freeway traffic by means of continuous-in-time microscopic 
models (car-following models) \cite{Reuschel}.
Since then, a multitude of car-following models have been
proposed, both for single-lane and multi-lane traffic including lane
changes.
We will restrict, here, to phenomena for which lane changes
are not important and only consider 
single-lane models.
To motivate our traffic model we first give an overview of the
dynamical properties of some popular microscopic models.

\subsection{Dynamic Properties of Some Car-Following Models}
%
Continuous-time single-lane car-following models are defined essentially by 
their acceleration function.
In many of the earlier models \cite{Chandler,Herman59,Gazis61,Edie},
the acceleration $\dot{v}_{\alpha}(t+T_r)$ of vehicle $\alpha$, delayed
by a reaction time $T_r$, can be written as
\begin{equation}
\label{gazis}
\dot{v}_{\alpha}(t+T_r) = \frac{- \lambda v_{\alpha}^m \Delta v_{\alpha}}
                            {s_{\alpha}^l}.
\end{equation}
The deceleration $-\dot{v}_{\alpha}(t+T_r)$ is asumed to be
proportional to the approaching rate
\begin{equation}
 \Delta v_{\alpha}(t) := v_{\alpha}(t)-v_{\alpha-1}(t)
\end{equation}
of vehicle
$\alpha$ {with respect} to the leading vehicle
$(\alpha-1)$.
In addition, the
acceleration may depend on the own velocity $v_{\alpha}$
\cite{Gazis61} and decrease with some power of
the net (bumper-to-bumper) distance
\begin{equation}
 s_{\alpha}=x_{\alpha-1}-x_{\alpha}-l_{\alpha}
\end{equation}
to the leading vehicle (where $l_{\alpha}$ is the vehicle length)
\cite{Gazis61,Edie}.
Since, according to Eq. (\ref{gazis}),
 the acceleration
depends on {a leading vehicle,}
these models are not applicable for very low traffic densities. If 
no leading {vehicle is} present (corresponding
to $s_{\alpha}\to\infty$), the acceleration is either
not determined $(l=0)$ or zero $(l>0)$, regardless of the own velocity.
However, one would expect in this case
that drivers accelerate to an individual desired
velocity.
The car-following behavior in dense traffic is also somewhat
unrealistic. In particular, the gap $s_{\alpha}$
to the respective front vehicle does not {necessarily} relax
to an equilibrium value. Even small gaps will not
induce braking reactions if the velocity difference 
$\Delta v_{\alpha}$ is zero.

These problems are solved by the car-following model 
of Newell \cite{Newell}. In this model, the velocity
at time $(t+T_r)$ depends adiabatically on the gap, i.e.,
the vehicle adapts exactly to a distance-dependent function $V$
within the reaction time $T_r$,
\begin{equation}
\label{newell}
v_{\alpha}(t+T_r) = V\Big(s_{\alpha}(t)\Big). 
\end{equation}
The ``optimal velocity function'' 
$V(s)=v_0\{1-\exp[-(s-s_0)/(v_0 T)]\}$ includes both, a desired velocity
$v_0$ for vanishing
interactions ($s\to\infty$) and a safe time headway $T$ 
characterizing the
car-following behavior in dense (equilibrium) traffic.
The Newell model is collision-free,
but the immediate dependence of the velocity
on the density leads to very high accelerations of the order of 
$v_0/T_r$. 
Assuming a typical desired velocity of
30 m/s and $T_r=1$ s, this would correspond to 30 m/s$^2$,  which is clearly
unrealistic \cite{Tilch-GFM,Bleile-Diss}. 

More than 30 years later, Bando {\it et al.} suggested a similar model,
\begin{equation}
\label{bando}
\dot{v}_{\alpha} = \frac{V(s_{\alpha}) - v_{\alpha}}{\tau}
\end{equation}
with a somewhat different optimal velocity function.
This ``optimal-velocity mdoel'' 
has been widely used by physicists because of its simplicity, and because
some results could be derived analytically.
The dynamical behavior 
does not  greatly differ from the Newell model,
since the reaction time delay $T_r$ of the Newell model can be compared
with the velocity relaxation time $\tau$ of the optimal-velocity  model.
However, realistic velocity relaxation
times are of the order of 10 s (city traffic) to 40 s 
(freeway traffic) and
therefore much larger than
reaction delay times (of the order of 1 s).
For typical values of the other
parameters of the optimal-velocity model  \cite{Bando},
crashes are only avoided if
$\tau < 0.9$ s, i.e., the velocity relaxation time is of the order
of the reaction time, leading again to
unrealistically high values 
$v_0/\tau$ 
of the maximum acceleration.
The reason of this unstable behavior is that {effects of}
velocity differences are neglected.
However, they play an essential stabilizing
role in real traffic, especially when approaching traffic jams. 
Moreover, in models (\ref{newell}) and (\ref{bando}),
accelerations and decelerations are symmetric with respect to the
deviation of the actual velocity {from} the equilibrium velocity,
which is unrealistic. The absolute value of
braking decelerations is usually stronger than that of
accelerations.

A relatively simple model with a generalized optimal velocity
function incorporating both, reactions to
velocity differences and
different rules for acceleration and braking has been proposed
rather recently \cite{Tilch-GFM}. This ``generalized-force model''
could  successfully reproduce
the time-dependent 
gaps and velocities measured by a sensor-equipped
car in congested city traffic. 
However, the acceleration and deceleration times in this model are
still unrealistically small which requires inefficiently small time
steps for the
numerical simulation.

Besides these simple models intended for basic investigations, 
there are also highly complex ``high-fidelity models'' 
like the Wiedemann model \cite{Wiedemann} or MITSIM \cite{MITSIM},
which try to reproduce traffic as realistically as possible, but
at the cost of a large number of parameters.

{Other} approaches that incorporate ``intelligent'' and
realistic braking reactions
are {the} simple and fast stochastic models
proposed by Gipps \cite{Gipps81,Gipps86} and Krauss 
\cite{Krauss-traff98}.
Despite their simplicity, these models show a  realistic
driver behavior, have asymmetric accelerations and decelerations,
and produce no accidents. Unfortunately,
they lose their realistic properties in the
deterministic limit. In particular, they show no traffic instabilities 
or hysteresis effects for vanishing fluctuations.

\subsection{Model Equations}

The acceleration assumed in the IDM is a continuous function 
of the velocity $v_{\alpha}$, the gap $s_{\alpha}$,
and the velocity difference (approaching rate)
$\Delta v_{\alpha}$ to the leading vehicle: 
\begin{equation}
\label{IDMv}
\dot{v}_{\alpha} = a^{(\alpha)}
         \left[ 1 -\left( \frac{v_{\alpha}}{v_0^{(\alpha)}} 
                  \right)^{\delta} 
                  -\left( \frac{s^*(v_{\alpha},\Delta v_{\alpha})}
                                {s_{\alpha}} \right)^2
         \right].
\end{equation}
This expression is {an interpolation of the tendency to accelerate 
with 
$a_f(v_{\alpha}) := a^{(\alpha)}[1-(v_{\alpha}/v_0^{(\alpha)})^{\delta}]$ 
on a free road and the tendency to brake with deceleration
$-b_{\rm int}(s_{\alpha}, v_{\alpha}, \Delta v_{\alpha})
:= -a^{(\alpha)}(s^*/s_{\alpha})^2$ when vehicle $\alpha$ comes too
close to the vehicle in front.} The deceleration term
depends on the ratio between the ``desired
minimum gap'' $s^*$ and the actual gap $s_\alpha$, where the desired gap
\begin{equation}
\label{sstar}
s^*(v, \Delta v) 
    = s_0^{(\alpha)} + s_1^{(\alpha)} \sqrt{\frac{v}{v_0^{(\alpha)}}}
    + T^{\alpha} v
    + \frac{v \Delta v }  {2\sqrt{a^{(\alpha)} b^{(\alpha)}}}
\end{equation}
%
%
is dynamically varying with the velocity and the approaching
rate.

In the rest of this paper, 
we will study the case of identical vehicles whose model
parameters $v_0^{(\alpha)}=v_0$, $s_0^{(\alpha)}=s_0$, 
$T^{(\alpha)}=T$, $a^{(\alpha)}=a$,
$b^{(\alpha)}=b$, and $\delta$ are
given
in Table \ref{tab:param}.
Here, our emphasis is on basic investigations with models as simple as
possible, and therefore
we will set $s_1^{\alpha}=0$ resulting in a model where
all parameters have an intuitive meaning with {plausible and often
easily measurable values.}
While the empirical data presented in this paper can be 
nevertheless reproduced, {a distinction of different driver-vehicle types
and/or} a nonzero $s_1$ \cite{TGF99-Treiber} is necessary
for a more quantitative agreement.
A nonzero $s_1$ would also be necessary for {features} 
requiring an inflection point in the equilibrium flow-density
relation, e.g., for certain types of
multi-scale expansions \cite{Nagatani-kink}.

\subsection{Dynamic Single-Vehicle Properties}
\begin{figure}[ht]
  \begin{center}
    \includegraphics[width=85mm]{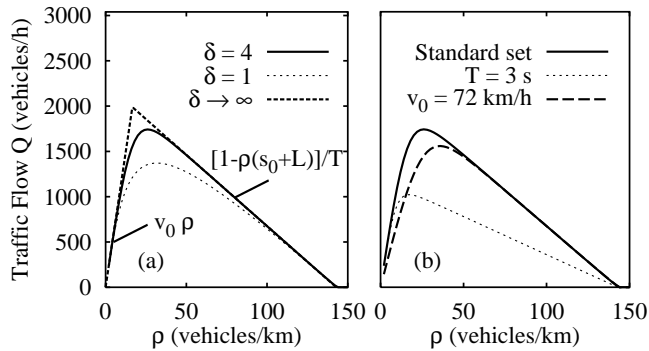}
  \end{center}
  \caption[]{\label{fig:fund}\protect 
Equilibrium flow-density relation of identical IDM 
vehicles with (a) variable
acceleration exponent $\delta$, and (b) variable
safe time headway $T$ and desired velocity $v_0$.
Only one parameter is varied at a time; the others
{correspond to the ones in} Table \ref{tab:param}.
}
\end{figure}

%
Special cases of the IDM acceleration (\ref{IDMv})
with $s_1=0$ include the following driving
modes:

\paragraph{Equilibrium traffic:}
In equilibrium traffic of arbitrary density
($\dot{v}_{\alpha}=0$, $\Delta v_{\alpha}=0$), drivers tend to keep 
a velocity-dependent
equilibrium gap $s_{e}(v_{\alpha})$ to the front vehicle
given by
\begin{equation}
\label{seq}
s_{e}(v) = s^*(v,0) 
  \left[ 1 - \left(\frac{v}{v_0}\right)^{\delta}\right]^{-\frac{1}{2}}
             = (s_0+v T) 
  \left[ 1 - \left(\frac{v}{v_0}\right)^{\delta}\right]^{-\frac{1}{2}}.
\end{equation}
In particular, the equilibrium gap of homogeneous 
{\it congested} traffic
({with} $v_{\alpha} \ll v_0$) is essentially equal
to the desired gap,
$s_{e}(v) \approx s_0+v T$, i.e., it is
composed of a bumper-to-bumper space  $s_0$ kept in
standing traffic and {an additional} velocity-dependent contribution
$v T$ corresponding to a constant safe time headway $T$.
This high-density limit is of
the same functional form as
that of the Newell model, Eq. (\ref{newell}).
Solving Eq. (\ref{seq}) 
for $v:=V_e(s)$ leads to simple expressions
only for $\delta=1$, $\delta=2$, or $\delta\to\infty$.
In particular, the equilibium velocity for
$\delta=1$ and $s_0=0$ is
\begin{equation}
\label{veGKT}
V_e(s)|_{\delta=1,s_0=0} = 
 \frac{s^2}{2 v_0 T^2} \left(-1 + \sqrt{1+\frac{4 T^2 v_0^2}{s^2}}
                       \right).
\end{equation}
Further interesting  cases  are
\begin{equation}
\label{vedelta2}
V_e(s)|_{\delta=2,s_0=0} = 
 \frac{v_0}{\sqrt{1+\frac{v_0^2 T^2}{s^2}}},
\end{equation}
and 
\begin{equation}
\label{vedeltainfty}
V_e(s)|_{\delta\to\infty} = 
\mbox{min} \{v_0, (s-s_0)/T \}.
\end{equation}
From a {\it macroscopic} point of view,
equilibrium traffic consisting of identical vehicles can
be characterized by the 
equilibrium traffic flow $Q_e(\rho)=\rho V_e(\rho)$ 
(vehicles per hour and per lane)
as a function of the
traffic density $\rho$ (vehicles per km and per lane). 
For the IDM model, this ``fundamental diagram'' 
follows from one of the equilibrium relations
(\ref{seq}) {to} (\ref{vedeltainfty}), together with the
micro-macro relation between gap and density:
\begin{equation}
\label{srho}
s=1/\rho-l = 1/\rho-1/\rho_{\rm max}.
\end{equation} 
Herein, the maximum density $\rho_{\rm max}$ is related
to the vehicle length $l$ by $\rho_{\rm max} l=1$.
Figure \ref{fig:fund} shows the fundamental diagram and its dependence
on the parameters
$\delta$, $v_0$, and $T$.
In particular, the fundamental diagram
for $s_0=0$ and $\delta=1$ is 
identical to the equilibrium relation
of the macroscopic GKT model, 
if the GKT parameter $\Delta A$ is set to zero 
(cf. Eq. (23) in Ref. \cite{GKT}), which is a necessary 
condition {for} a micro-macro {correspondence} \cite{TGF99-Hennecke}.

\begin{figure}[ht]
  \begin{center}
    \includegraphics[width=85mm]{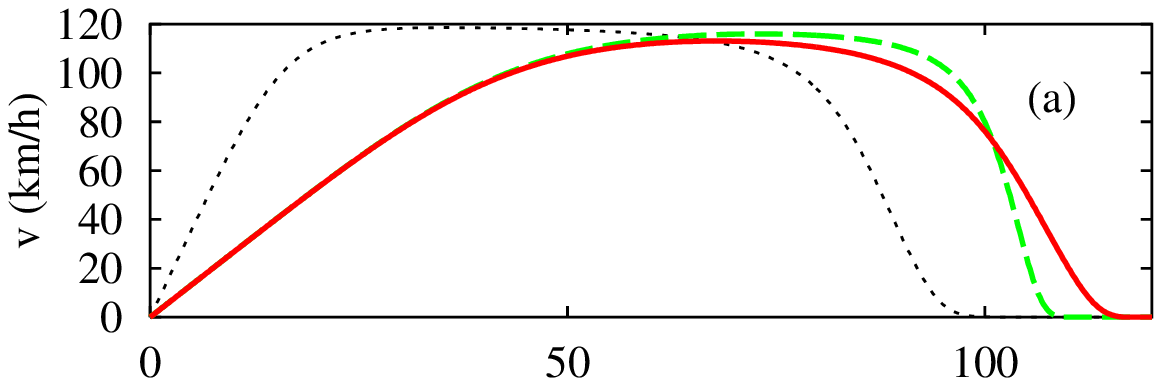}\\[-3mm]
    \includegraphics[width=85mm]{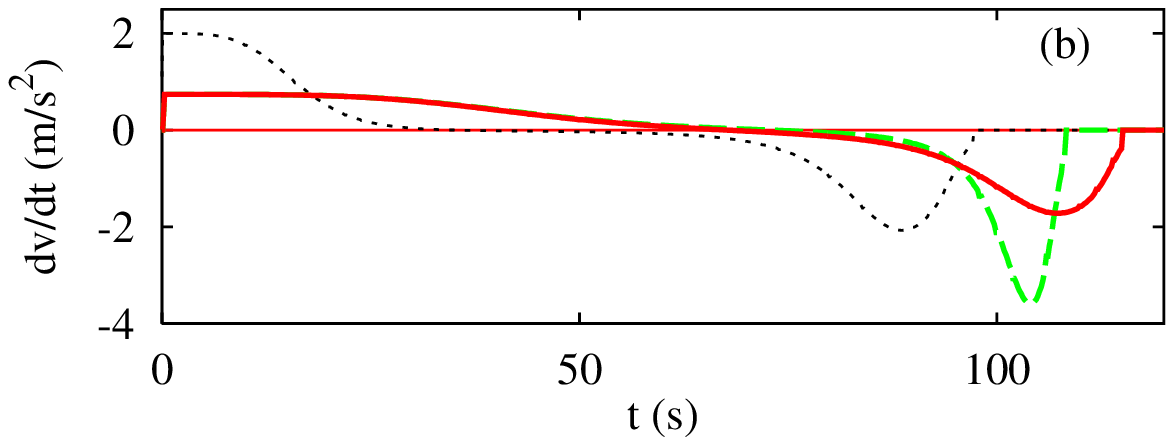}
  \end{center}
  \caption[]{\label{fig:accbrake}\protect 
Temporal evolution of velocity and 
acceleration of a single driver-vehicle unit
which accelerates on a 2.5 km long stretch of free road
before it decelerates when approaching 
a standing obstacle at $x=2.5$ km.
The dynamics for the IDM parameters of Table \ref{tab:param}
(solid) is compared with the result for an increased
acceleration $a_0=2$ m/s (dotted), or an increased braking
deceleration $b=5$ m/s$^2$ (dashed).
}
\end{figure}

\paragraph{Acceleration to the desired velocity:}
If the traffic density is very low
($s$ is large), the interaction term is negligible
and the IDM acceleration reduces to the free-road acceleration
$a_f(v)=a(1-v/v_0)^{\delta}$, which is a decreasing
function of the velocity with a maximum value
$a_f(0)=a$ and $a_f(v_0)=0$.
In Fig. \ref{fig:accbrake}, this regime applies for times
$t\le 60$ s.
The acceleration exponent $\delta$ specifies
how the
acceleration decreases when approaching the desired velocity.
The limiting case $\delta\to\infty$ corresponds to approaching
$v_0$ {with} a constant acceleration $a$, while $\delta=1$
corresponds to an exponential relaxation to the desired velocity 
with the {relaxation time} $\tau=v_0/a$.
In the latter case, the free-traffic acceleration is equivalent to that of
the optimal-velocity model (\ref{bando})
and also to acceleration functions of many
macroscopic models like the KKKL model
\cite{KK-94}, or the 
GKT model \cite{GKT}.
However, {the most realistic behavior is expected in between the
two limiting cases of exponential acceleration (for $\delta = 1$) and constant
acceleration (for $\delta \rightarrow \infty$), which is confirmed by
our simulations with the IDM. Throughout this paper we will use
$\delta = 4$.}

\paragraph{Braking as reaction  to high approaching rates:} 
When
approaching slower or standing vehicles with 
sufficiently high approaching rates $\Delta v>0$,
the equilibrium part $s_0+v T$ of the dynamical desired
distance $s^*$, Eq. (\ref{sstar}),
can be neglected with respect to the
nonequilibrium 
part, which is proportional to $v\Delta v$.
Then, the interaction part $-a(s^*/s)^2$
of the acceleration equation (\ref{IDMv})
is given by
\begin{equation}
\label{aapproach}
b_{\rm int}(s, v, \Delta v)
  \approx \frac{(v\Delta v)^2}{4bs^2}.
\end{equation}
This expression implements anticipative ``intelligent''
braking behavior, which we disuss now for the spacial case of
approaching a standing obstacle
($\Delta v=v$).
Anticipating a constant deceleration during the whole approaching
process,
a minimum kinematic deceleration
$b_k := v^2/(2 s)$ is
necessary to avoid a collision.
The situation is assumed to be ``under control'', if $b_k$
is smaller than the 
``comfortable'' deceleration given by the model parameter $b$,
i.e., $\beta := b_k/b \le 1$. In contrast, an emergency
situation is characterized by $\beta>1$.
With these definitions, Eq. (\ref{aapproach}) becomes
\begin{equation}
\label{aaproach1}
b_{\rm int} (s,v,v) = \frac{b_k^2}{b} = \beta b_k. 
\end{equation}
While in safe situations the IDM deceleration
is less than the
kinematic collision-free deceleration,
drivers overreact in
emergency situations to get the
situation again under control. 
It is easy to show that in both cases 
the acceleration
approaches $\dot{v}=-b$ under the deceleration law (\ref{aaproach1}).
Notice that this stabilizing behavior is lost if one replaces in
Eq. (\ref{IDMv}) the braking term $-a(s^*/s)^2$ by
$-a'(s^*/s)^{\delta'}$ with $\delta' \le 1$
corresponding to $b_{\rm int}(s,v,v)=\beta^{\delta'-1} b_k$.
%
The ``intelligent'' braking behavior of drivers in
this regime makes the model collision-free.
The right parts of the plots of Fig. \ref{fig:accbrake} 
($t>70$ s) show the simulated approach of an IDM vehicle to 
a standing obstacle.
As expected, the maximum deceleration is of the order of $b$.
For low velocities, however, the equilibrium term $s_0+vT$ of $s^*$
cannot be neglected as assumed when deriving Eq. (\ref{aaproach1}).
Therefore, the maximum deceleration is somewhat lower than $b$ and the 
deceleration decreases immediately before the stop while,
under the dynamics (\ref{aaproach1}), one would have $\dot{v}=-b$.

Similar braking rules have  been implemented in the model
of Krauss \cite{Krauss-traff98},
where the model
is formulated in terms of a time-discretizised update scheme 
(iterated map), where the velocity at timestep $(t+1)$ 
is limited to a
``safe velocity'' which is calculated on the basis of
the kinematic braking distance at a given ``comfortable'' deceleration.

\begin{figure}[ht]
  \begin{center}
    \includegraphics[width=85mm]{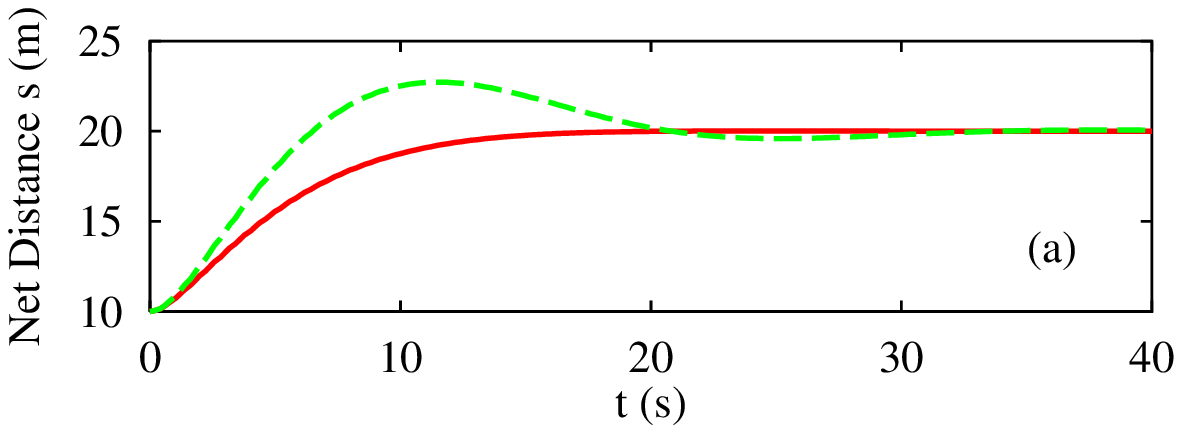}\\[-1mm]
    \includegraphics[width=85mm]{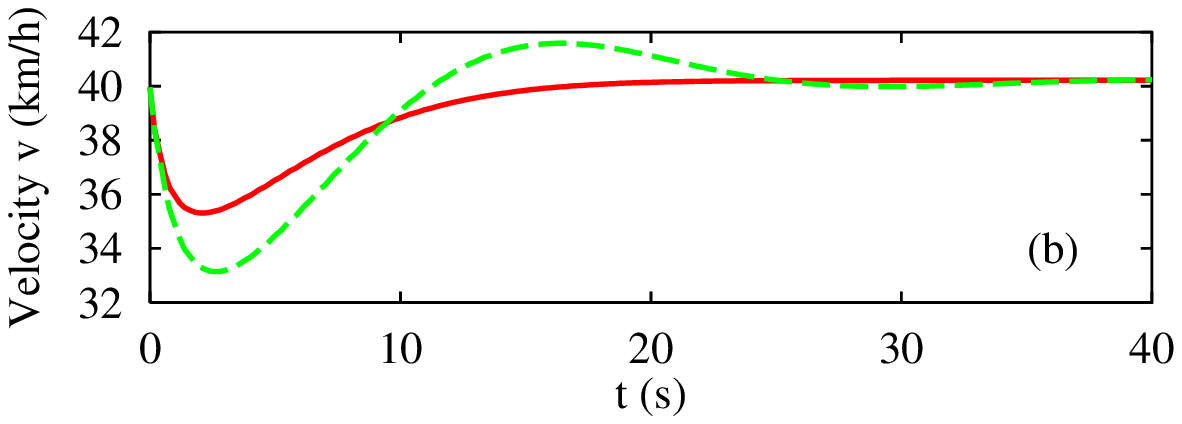}
  \end{center}
  \caption[]{\label{fig:distance}\protect 
Adaptation of a single vehicle 
to the equilibrium distance in the car-following regime.
Shown is (a) the net distance $s$, and (b) the velocity of a 
vehicle following
a queue of vehicles which all drive at $v^*=40.5$ km/h corresponding
to an equilibrium distance $s_{e}=20$ m.
The initial conditions are $v(0)=v^*$ and $s(0)=s_{e}/2=10$ m.
The solid line  is for the IDM standard parameters,
and the dashed line for the deceleration parameter $b$ increased 
from 1.67 m/s$^2$ to 10 m/s$^2$.
}
\end{figure}

\paragraph{Braking in response to small gaps:}
The forth driving mode is active when 
the gap is much smaller than $s^*$ but there are no large velocity
differences.
Then, the equilibrium part
$s_0+v T$ of $s^*$ dominates over the dynamic contribution
proportional to $\Delta v$. Neglecting the 
free-road acceleration, Eq. (\ref{IDMv})
reduces to 
$\dot{v} 
\approx -(s_0+v T)^2/s^2$,
corresponding
to a Coulomb-like repulsion. Such braking
interactions are also implemented
in other models, e.g., in
the model of Edie \cite{Edie}, the GKT model \cite{GKT},
or in certain regimes of the Wiedemann
model \cite{Wiedemann}.
The dynamics in this driving regime is not qualitatively different,
if one replaces $-a(s^*/s)^2$ by
$-a(s^*/s)^{\delta'}$ with $\delta' >0$. This is in contrast to the
approaching regime, where collisions would be 
{provoked} for $\delta' \le 1$.
Figure \ref{fig:distance} shows the car-following dynamics in this
regime.
For the standard parameters, one clearly sees an 
{non-oscillatory} relaxation
to the equilibrium distance {(solid curve), 
while for very high values of $b$, the approach to the 
equilibrium distance would occur 
with damped oscillations (dashed curve).}
Notice that, for the latter parameter
set, the {\it collective} {traffic dynamics}
would already be extremely unstable.

\subsection{Collective Behavior and Stability Diagram}

\begin{figure}[ht]
  \begin{center}
    \includegraphics[width=85mm]{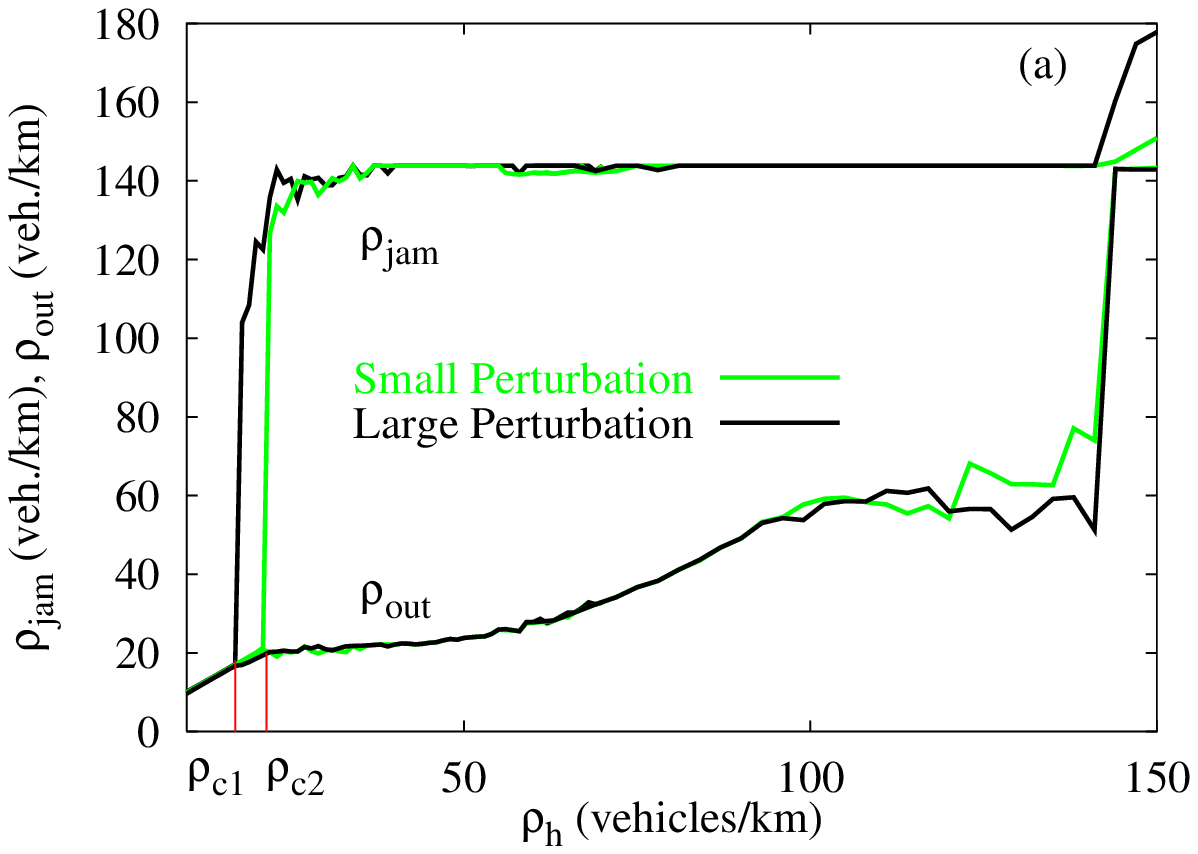}\\
    \includegraphics[width=85mm]{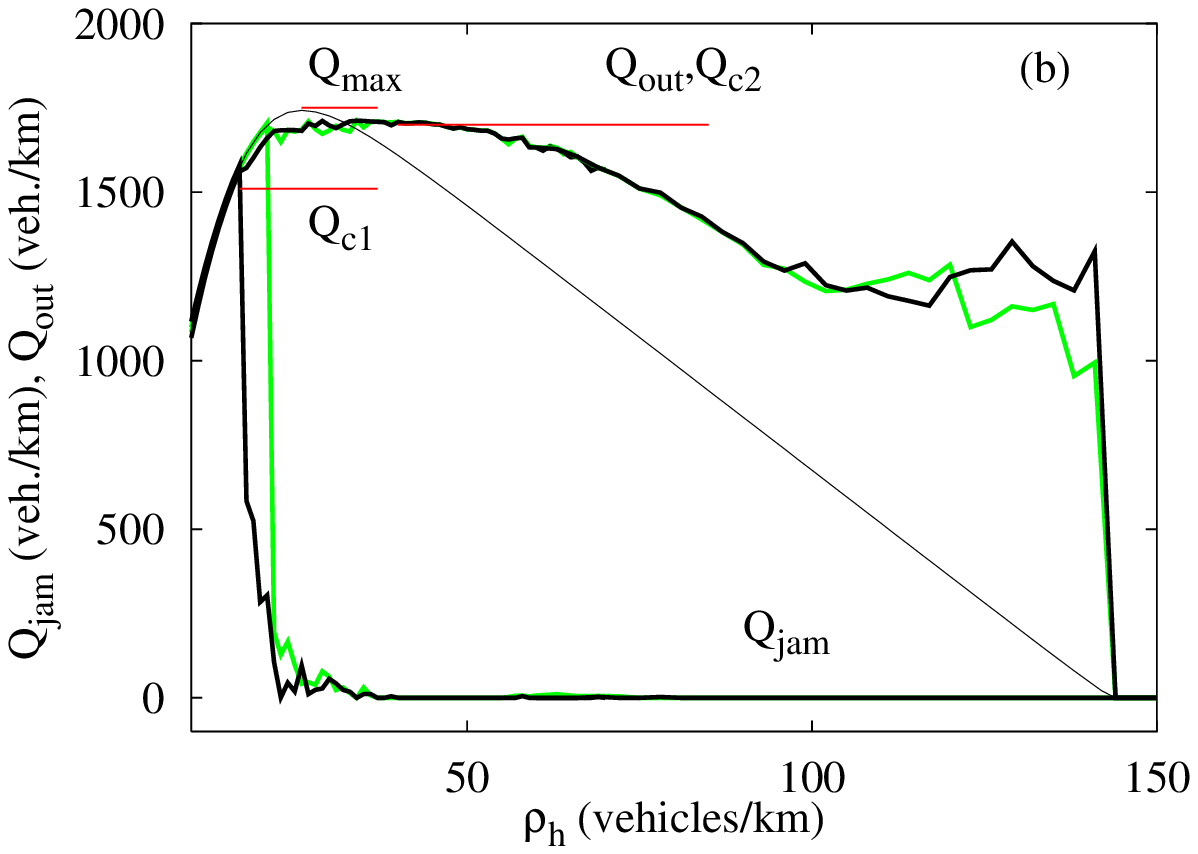}
  \end{center}
  \caption[]{\label{fig:stab}\protect 
Stability diagram of homogeneous traffic 
{on a circular road}
as a function of the homogeneous density $\rho_h$
for small (grey) and large (black) initial
perturbations of the density. In plot (a),
the upper two lines display the density inside
of density clusters after a stationary state has been reached. 
The lower two lines represent the density between the clusters.
Plot (b) shows the corresponding flows and the
equilibrium flow-density relation (thin curve).
The critical densities $\rho_{{\rm c}i}$ and flows
$Q_{{\rm c}i}$ are discussed in the 
main text. For $\rho_{\rm c2}\le\rho_h\le 45$ vehicles/km, the outflow
$Q_{\rm out}\approx Q_{\rm c2}$ and the corresponding density
$\rho_{\rm out}$ are constant. Here, we have
$Q_{\rm max}\approx Q_{\rm out}$ 
for the maximum equilibrium flow, but there are other parameter sets
(especially if $s_1>0$) where $Q_{\rm max}$ is clearly
larger than $Q_{\rm out}$ \protect\cite{TGF99-Treiber}.
}
\end{figure}

Although we are interested in realistic 
{\it open} {traffic} systems, 
it turned out that many features can be explained
in terms of the stability behavior in a {\it closed} system.
Figure \ref{fig:stab}(a) shows the stability diagram of 
homogeneous traffic
on a circular road. The control parameter is the
homogeneous density $\rho_{\rm h}$. We applied both a very small 
and a large localized perturbation to check for linear and nonlinear
stability, and plotted the resulting minimum
($\rho_{\rm out}$) and maximum ($\rho_{\rm jam}$)
densities after a stationary situation was reached. 
The resulting diagram is very similar to that of the
macroscopic KKKL and GKT models \cite{KK-94,GKT}. 
In particular, it displays the following realistic features:
(i) Traffic is stable for very low and high densities, but unstable for
intermediate densities. 
(ii) There is a density range $\rho_{\rm c1}\le\rho_{\rm h}\le\rho_{\rm c2}$
of metastability, i.e., only perturbations of sufficiently large
amplitudes grow, while smaller perturbations disappear. Note that, 
for most IDM parameter sets, there is no second metastable range at 
higher densities, in contrast to the GKT and KKKL models. 
Rather, traffic flow becomes stable again for densities exceeding the
critical density $\rho_{\rm c3}$, or {congested} flows below
$Q_{\rm c3}=Q_{\rm e}(\rho_{\rm c3})$.
(iii) The density $\rho_{\rm jam}$ inside of traffic jams 
and the associated flow $Q_{\rm jam}=Q_{\rm e}(\rho_{\rm jam})$,
cf. Fig. \ref{fig:stab}(b),
do not depend on
$\rho_{\rm h}$. For the parameter set chosen here, we have
$\rho_{\rm jam}=\rho_{\rm c3}=140$ vehicles/km, and $Q_{\rm jam}=0$,
{i.e., there is no linearly stable congested traffic with a finite
flow and velocity.}
For other parameters, especially for a nonzero
IDM parameter $s_1$, both $Q_{\rm jam}$ and $Q_{\rm c3}$ 
can be nonzero and
different from each other \cite{TGF99-Treiber}.

As further ``traffic constants'', at least
in the density range 20 veh./km $\le \rho_{\rm h} \le$ 50 veh./km, 
we observe a constant outflow $Q_{\rm out}=Q_{\rm e}(\rho_{\rm out})$
and propagation velocity 
$v_{\rm g}=(Q_{\rm out} - Q_{\rm jam}) / (\rho_{\rm out}-\rho_{\rm jam})
\approx -15$ km/h of {traffic} jams.
Figure \ref{fig:stab}(b) shows the stability diagram for the flows.
In particular, 
we have $Q_{\rm c1}<Q_{\rm out} \approx Q_{\rm c2}$,
where $Q_{{\rm c}i}=Q_{\rm e}(\rho_{{\rm c}i})$, i.e., the outflow from
congested traffic is at the margin of linear stability, which is
also the case in the GKT for most parameter sets \cite{GKT,Phase}.
For other IDM parameters, the outflow $Q_{\rm out}\in
[Q_{\rm c1},Q_{\rm c2}]$ is metastable \cite{TGF99-Treiber}, or even at the 
margin of nonlinear stability \cite{coexist}.
\par
In {\it open} systems, a third type of stability becomes relevant.
Traffic is {\it convectively} stable, if, after a sufficiently long time,
all perturbations are convected out of the system.
Both, in the macroscopic models and in the IDM, there is a considerable
density region $\rho_{\rm cv}\le \rho_{\rm h}\le \rho_{\rm c3}$,
where traffic is linearly unstable but convectively stable.
{For the parameters chosen in this paper,} congested traffic is 
{\it always} linearly unstable, but convectively stable for flows below 
$Q_{\rm cv}=Q_{\rm e}(\rho_{\rm cv})=1050$ vehicles/h.
A nonzero jam distance $s_1$ is required for linearly
{\it stable} congested traffic with nonzero flows 
\cite{TGF99-Treiber}, at least, if the model 
should simultaneously show traffic instabilities.
\subsection{Calibration}
Besides the vehicle length $l$, the IDM has seven parameters,
cf. Table \ref{tab:param}.
The {\it fundamental relations}
of homogeneous traffic are calibrated
with the desired velocity $v_0$ (low density), 
safe time headway $T$ (high density), and the jam distances
$s_0$ and $s_1$ (jammed traffic).
In the low-density limit $\rho\ll (v_0 T)^{-1}$, 
the equilibrium flow can be approximated by
$Q_e\approx V_0\rho$. In the high density regime and for $s_1=0$,
one has a linear decrease of the flow 
$Q_e\approx [1-\rho(l+s_0)]/T$ which can be used to determine
$(l+s_0)$ and $T$. 
Only for nonzero $s_1$, one obtains an inflection point
in the equilibrium flow-density relation $Q_e(\rho)$.
The acceleration coefficient $\delta$ 
influences the transition region between the free and
congested regimes. For $\delta\to\infty$ and $s_1=0$, 
the fundamental diagram (equilibrium flow-density relation) becomes
triangular-shaped: $Q_e(\rho) = \mbox{min}(v_0\rho,
[1-\rho(l+s_0)]/T)$. 
For decreasing $\delta$, it becomes smoother and 
smoother, cf. Fig. \ref{fig:fund}(a).

The {\it stability behavior} of traffic in the IDM model
is determined mainly by
the maximum acceleration $a$, desired deceleration $b$, 
and by $T$. Since the accelerations $a$ and $b$ 
do not influence the fundamental diagram,
the model can be calibrated essentially independently with respect to
traffic flows and stability.
As in the GKT model, traffic becomes more unstable for
decreasing $a$ (which corresponds to an increased acceleration time
$\tau=v_0/a$), and with decreasing $T$
(corresponding to reduced safe time headways). 
Furthermore, the instability increases with 
increasing $b$. This is also plausible, because an increased 
desired deceleration $b$
corresponds to a less anticipative or less defensive braking
behavior.
The density and flow in jammed traffic
and the outflow from traffic jams is also influenced by $s_0$ and $s_1$.
In particular, for $s_1=0$, the traffic flow $Q_{\rm jam}$
inside of traffic jams 
is typically zero after a sufficiently long time [Fig. \ref{fig:stab}(b)],
but nonzero otherwise.
The stability of the self-organized outflow $Q_{\rm out}$ depends
strongly on the minimum jam distance $s_0$. It can be unstable
(small $s_0$), metastable, 
or stable (large $s_0$). In the latter
case, traffic instabilities can only lead to single localized
clusters, not to stop-and go traffic.

\section{\label{sec:simu} Microscopic Simulation of 
Open Systems with an Inhomogeneity}

We simulated identical vehicles of length $l=5$ m
{with the typical} IDM model parameters listed in
Table~\ref{tab:param}.  
{Moreover,} although the various congested states discussed 
in the following were
observed on different freeways, {all of them were qualitatively reproduced
with very restrictive variations of one single parameter (the safe
time headway $T$), while we always used the same values for the other
parameters (see Table~\ref{tab:param}).} 
This indicates that the model is quite realistic and
robust. Notice that all parameters have plausible values.
The value $T=1.6$ s for the safe time headway 
is slightly lower than suggested by
German authorities (1.8 s).
The acceleration parameter $a=0.73$ m/s$^2$ corresponds to
a free-road acceleration from $v=0$ to $v=100$ km/h within 45 s,
cf. Fig. \ref{fig:accbrake}(b).
This value is obtained by 
integrating the IDM acceleration 
$\dot{v}=a[1-(v/v_0)^{\delta}]$ with  
$v_0=120$ km/h, $\delta=4$, and the initial value $v(0)=0$. 
While this is considerably above
minimum acceleration times 
(10 s - 20 s for average-powered cars),
it should be characteristic for {\it everyday} accelerations. 
The comfortable deceleration 
$b=1.67$ m/s$^2$ is also consistent with empirical investigations
\cite{Bleile-Diss,Tilch-GFM}, 
and with parameters used in more complex models \cite{MITSIM}.

With efficient numerical integration schemes,
we obtained
a numerical performance of about $10^5$
vehicles in realtime on a usual workstation \cite{note-phase2-micperf}.

\subsection{Modelling of Inhomogeneities}
%
Road inhomogeneities can be classified into flow-conserving
local defects like narrow road sections or gradients, 
and those that do not conserve the average flow per lane,
like on-ramps, off-ramps, or lane closings. 

{\it Non-conserving} inhomogeneities can be incorporated 
into macroscopic models in a natural way by 
adding a source term to the continuity equation 
for the {vehicle} density \cite{Kerner-ramp,sync-Letter,Lee}.
An explicit microscopic modelling of on-ramps or lane closings,
however, would require
a multi-lane model with an explicit simulation of lane changes. 
Another possibility {opened by the recently formulated micro-macro link
\cite{TGF99-Hennecke} is to} 
simulate the ramp section {macroscopically with a} source term in
the continuity equation \cite{sync-Letter}, {and to simulate the
remaining stretch microscopically.}

In contrast, {\it flow conserving} inhomogeneities
can be implemented easily in both microscopic and macroscopic
single-lane models by locally changing the values of one or more
model parameters or by imposing external decelerations
\cite{Nagatani-sync}.
Suitable parameters for the IDM and the GKT model
are the desired velocity $v_0$, or the safe time headway $T$.
Regions with locally decreased desired velocity
can be interpreted
either as sections with speed limits, or as sections with 
uphill 
gradients (limiting the maximum 
velocity of some vehicles) \cite{note-phase2-speedlimit}.
Increased safe time headways can be
attributed to more careful driving behavior {along curves,} on
narrow, dangerous road sections, {or a reduced range of visibility}.

\begin{figure}[ht]
  \begin{center}
    \includegraphics[width=85mm]{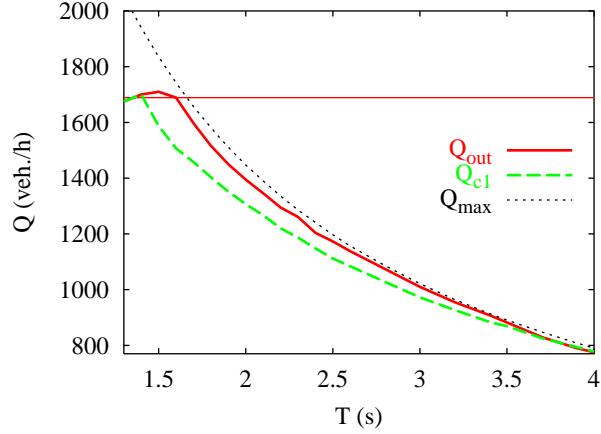}
  \end{center}
  \caption[]{\label{fig:Qout}\protect 
 Outflow $Q_{\rm out}$ from congested traffic (solid curve), 
minimum flow $Q_{\rm c1}$ for nonlinear instabilities (dashed),
and maximum equilibrium flow $Q_{\rm max}$ (dotted)
as a function of the safe time headway.
An approximation for the bottleneck strength 
$\delta Q$ of the phase diagram
is given by the difference between the value 
$Q_{\rm out}=1689$ vehicles/h for $T=1.6$ s (horizontal thin line),
and $Q_{\rm out}(T)$. For decerasing values of $T$ 
traffic becomes more unstable which is indicated by increasing
differences ($Q_{\rm max} - Q_{\rm out}$) or 
            ($Q_{\rm max} - Q_{\rm c1}$).
For $T\le 1.5$ s this even leads to
$\partial Q_{\rm out}(T)/\partial T > 0$. Furthermore,
$Q_{\rm out} \approx Q_{\rm c1}$ for
$T\le 1.4$ s.
}
\end{figure}

Local parameter variations act as a bottleneck,
if the outflow $Q'_{\rm out}$ from congested traffic 
(dynamic capacity) in the downstream
section is reduced with
respect to the outflow $Q_{\rm out}$ in the upstream section.
This outflow 
can be determined from fully developed
stop-and go waves in a {\it closed} 
system, whose outflow is constant in a rather large range of 
average densities $\rho_{\rm h}\in$ [20 veh./km, 60 veh./km],
cf. Fig. \ref{fig:stab}(b). 
It will turn out that the outflow $Q'_{\rm out}$ is the relevant
capacity for understanding congested traffic, 
and not the maximum flow $Q_{\rm max}$ ({\rm static capacity}),
which can be reached in (spatially homogeneous) equilibrium traffic only.
The capacities are decreased, e.g., for 
a reduced desired velocity $v'_0<v_0$
or an increased safe time headway $T'>T$, or both.
Figure \ref{fig:Qout} shows $Q_{\rm out}$ and $Q_{\rm max}$
as a function of $T$.
For $T'>4$ s,
traffic flow is always
stable, and the outflow from jams is equal to the static capacity.

For {\it extended} congested states, 
all types of flow-conserving bottlenecks
result in a similar traffic dynamics, if
$\delta Q = (Q_{\rm out} - Q'_{\rm out})$ is identical,
where $Q'_{\rm out}$ is the outflow for the changed model parameters
$v_0'$, $T'$, etc. 
Qualitatively the same dynamics is also observed in 
{\it macroscopic} models including
on-ramps, if the ramp flow satisfies
$Q_{\rm rmp}\approx \delta Q$ \cite{TGF99-Treiber}.
This suggests to introduce the following general definition of
the ``bottleneck strength'' $\delta Q$:
\begin{equation}
\label{deltaQ}
\delta Q := Q_{\rm rmp} + Q_{\rm out}-Q'_{\rm out}, 
\end{equation}
In particular, we have $\delta Q=Q_{\rm rmp}$ for on-ramp bottlenecks,
and $\delta Q=(Q_{\rm out}-Q'_{\rm out})$ for flow-conserving
bottlenecks, {but formula (\ref{deltaQ}) is also applicable for a
combination of both.}
\subsection{Phase Diagram of Traffic States in Open Systems}
%
In contrast to {\it closed} systems, in which the long-term behavior and
stability is essentially
determined by the average traffic density $\rho_h$, the dynamics of 
{\it open} systems
is controlled by the inflow $Q_{\rm in}$ {to the main road (i.e., the
flow at the upstream boundary).}
Furthermore, traffic congestions depend on road inhomogeneities and,
because of hysteresis effects, on the history of
previous perturbations.

In this paper, we will implement flow-conserving
inhomogeneities by a variable safe time headway
$T(x)$. We chose  $T$ as variable model parameter because it
influences the flows {more effectively than} $v_0$,
which has been {varied} in Ref. \cite{TGF99-Treiber}.
Specifically, we increase the
local safe time headway according to
\begin{equation}
\label{Tx} 
T(x)=\left\{ \begin{array}{ll}
  T  & \ \ x\le -L/2 \\
  T' & \ \ x\ge  L/2 \\
  T + (T'-T) \left(\frac{x}{L} + \frac{1}{2} \right)
     & \ \ |x|<L/2,
\end{array} \right.
\end{equation}
where the transition region of length
$L=600$ {is analogous to the ramp length for inhomogeneities that do
not conserve the flow.} 
The bottleneck strength $\delta Q(T')=[Q_{\rm out}(T)-Q_{\rm out}(T')]$
is an increasing function of $T'$, cf. Fig. \ref{fig:Qout}(a).
We investigated the traffic dynamics for various points
$(Q_{\rm in}, \delta Q)$ or, alternatively,
$(Q_{\rm in}, T' - T)$ 
in the control-parameter space.

Due to hysteresis effects and multistability, the phase diagram,
i.e., the asymptotic traffic state as a function of the control parameters
$Q_{\rm in}$ and $\Delta T$, depends also on the {\it history}, i.e.,
on initial conditions and on past boundary conditions and
perturbations.
Since we cannot explore the whole functional space of initial 
conditions, boundary conditions, and past perturbations, 
we used the following three representative ``standard'' 
histories.
%
\begin{itemize}
\item [A.] Assuming very low values for the initial density and flow,
we slowly increased the inflow to the prescribed value
$Q_{\rm in}$. 
\item [B.] 
We started the simulation with a stable pinned localized cluster
(PLC) state and a
consistent value for the inflow $Q_{\rm in}$. Then, we 
adiabatically changed the inflow
to the values prescribed by the point in the phase diagram.
\item [C.] After running history A, we 
applied a large perturbation at the downstream boundary.
If traffic at the given phase point is metastable, this initiates
an upstream propagating localized cluster which finally crosses
the inhomogeneity, {see} Fig.~\ref{3dphases}(a)-(c).
If traffic is unstable, the breakdown already occurs {during time
period A,} and the additional perturbation has no dynamic influence.
\end{itemize}
%
For a given history, the resulting phase diagram is unique.
The solid lines of Fig. \ref{phasediag} 
show the IDM phase diagram for History {C}.
Spatio-temporal density 
plots of the congested traffic states themselfes are displayed in
Fig.~\ref{3dphases}. To obtain the spatiotemporal density $\rho(x,t)$
from the microscopic quantities,
we generalize the micro-macro relation (\ref{srho}) to
define the density at discrete positions 
$x_{\alpha} + \frac{l+s_{\alpha}}{2}$ centered
between vehicle $\alpha$ and its predecessor,
\begin{equation}
\label{rhoxt_s}
\rho(x_{\alpha} + \frac{l+s_{\alpha}}{2}) = \frac{1}{l+s_{\alpha}},
\end{equation}
and interpolate linearly between these positions.
Depending on $Q_{\rm in}$ and $\delta T:=(T'-T)$, the downstream perturbation
(i) dissipates, resulting in free traffic (FT), 
(ii) travels through the
inhomogeneity as a moving localized cluster (MLC)
and neither dissipates nor triggers new breakdowns,
(iii) triggers a traffic
breakdown to a pinned localized cluster (PLC), which remains 
localized near the inhomogeneity for 
all times and either is stationary (SPLC),
cf. Fig.~\protect\ref{tristab}(a) for $t<0.2$ h,
or  oscillatory (OPLC).
(iv) Finally, the initial perturbation can induce 
extended congested traffic (CT), whose downstream boundaries
are fixed at the inhomogeneity, while the upstream front propagates
further upstream in the course of time. Extended congested traffic 
can be homogeneous (HCT), oscillatory (OCT), or consist of
triggered stop-and-go waves (TSG). We also include in the HCT region a 
complex state (HCT/OCT) where traffic is homogeneous only near the bottleneck,
but growing oscillations develop further upstream.
In contrast to OCT, where there is
permanently congested traffic at the inhomogeneity (``pinch region''
\cite{Kerner-wide,coexist}), the TSG state is characterized by a
series of isolated density clusters, each of which triggers a new
cluster as it passes the inhomogeneity. 
The maximum perturbation of History C, used also in Ref. \cite{Phase}, 
seems to select always the
stable extended congested phase.
We also scanned the control-parameter space $(Q_{\rm in},\delta Q)$
with Histories A and B exploring the maximum phase space of the 
(meta)stable FT and PLC states, respectively, cf. Fig. \ref{phasediag}.
In multistable regions of the control-parameter
space, the three histories can be used to select
the different traffic states, see below.

\subsection{Multistability}
In general, the phase
transitions between free traffic,
pinned localized states, and extended congested states are hysteretic. 
In particular, in all four examples of Fig. \ref{3dphases},
free traffic is possible as a second, metastable state. 
In the regions between the two dotted lines of the
phase diagram Fig.~\ref{phasediag}, both, free and congested traffic is 
possible, depending on the previous history. 
In particular, 
for all five indicated phase points, 
free traffic would persist without the downstream perturbation. In contrast,
the transitions PLC-OPLC, and
HCT-OCT-TSG seem to be non-hysteretic, i.e., the type of pinned
localized cluster or of extended congested traffic, is uniquely
determined by $Q_{\rm in}$ and $\delta Q$.

In a small subset of the metastable region, labelled ``TRI'' in
Fig. \ref{phasediag}, we even found {\it tristability} {with the
possible states} FT, PLC, and OCT. 
Figure \ref{tristab}(a) shows that
a single moving localized cluster passing the inhomogeneity
triggers a transition from PLC to OCT. Starting with free traffic,
the same perturbation would trigger OCT as well
[Fig. \ref{tristab}(b)], while we never found reverse
transitions OCT $\to$ PLC or OCT $\to$ FT (without a reduction of the 
inflow). That is, FT and PLC are metastable in the tristable region, 
while OCT is stable.
We obtained qualitatively the same also for the 
macroscopic GKT
model with an on-ramp as inhomogeneity [Fig. \ref{tristab}(c)].
Furthermore, tristability between FT, OPLC, and OCT
has been found for the IDM model with variable $v_0$ \cite{TGF99-Treiber},
and for the KKKL model \cite{Lee99}.

{Such a tristability can only exist if
the (self-organized) outflow 
$Q_{\rm out}^{\rm OCT}= \tilde{Q}_{\rm out}$
from the OCT state is lower than
the maximum outflow $Q_{\rm out}^{\rm PLC}$ from the PLC state.
A phenomenological explanation of this condition can be inferred from
the positions of the downstream fronts of the OCT and PLC states
shown in Fig. \ref{tristab}(a). The downstream front
of  the OCT state ($t>1$ h) is at $x\approx 300$ m, i.e., at the
{\it downstream boundary} of the $L=600$ m wide transition region, 
in which the
safe time headway Eq. (\ref{Tx}) increases from $T$ to $T'>T$.
Therefore, the local safe time headway
at the downstream front of OCT is $T'$, or
$Q_{\rm out}^{\rm OCT} \approx Q'_{\rm out}$, which was also used to
derive Eq. (\ref{Qcong}).
In contrast, the PLC state is centered at about $x=0$, so that
an estimate for the upper boundary $Q_{\rm out}^{\rm PLC}$
of the outflow is given by the self-organized outflow $Q_{\rm out}$ 
corresponding to
the local value $T(x)=(T+T')/2$ of the safe time headway at $x=0$.
Since $(T+T')/2 < T'$ this outflow is higher than 
$Q_{\rm out}^{\rm OCT}$ (cf. Fig. \ref{fig:Qout}).
It is an open question, however, why the downstream front of the OCT
state is further downstream compared to the PLC state.
Possibly, it can be explained by the close relationship of OCT with the TSG
state, for which the newly triggered density clusters even enter the
region downstream of the bottleneck
[cf. Fig. \ref{3dphases}(c)]. In accordance with its relative
location in the phase diagram, it is plausible that the OCT state
has a ``penetration depth'' into the downstream area that is
in between the one of the PLC and the TSG states.
}
%
%
%
\subsection{Boundaries between and Coexistence of Traffic States}
Simulations show that the outflow $\tilde{Q}_{\rm out}$
from the nearly stationary downstream fronts of OCT and HCT
satisfies $\tilde{Q}_{\rm out}\le Q'_{\rm out}$, where
$Q'_{\rm out}$ is the outflow from {fully developed density}
clusters in homogeneous systems for
the downstream model parameters. If the bottleneck is not too strong
(in the phase diagram Fig. \ref{phasediag}, it must satisfy
$\delta Q<350$ vehicles/h),
we have $\tilde{Q}_{\rm out}\approx Q'_{\rm out}$.
Then, for all types of bottlenecks, the congested traffic flow is
given by
$Q_{\rm cong}=\tilde{Q}_{\rm out} - Q_{\rm rmp} \approx 
Q'_{\rm out}  - Q_{\rm rmp}$, or
\begin{equation}
\label{Qcong}
Q_{\rm cong} 
\approx Q_{\rm out} - \delta Q.
\end{equation}
Extended congested traffic (CT) 
only persists, if the inflow $Q_{\rm in}$ exceeds the congested
traffic flow $Q_{\rm cong}$. Otherwise, it dissolves to PLC.
This gives the boundary
\begin{equation}
\label{CT-PLC}
\mbox{CT}\to\mbox{PLC}: \delta Q\approx Q_{\rm out}-Q_{\rm in}.
\end{equation}
If the traffic flow of CT states is 
{\it linearly} stable (i.e., 
$Q_{\rm cong} < Q_{\rm c3}$), we have HCT. If, for higher flows, it is
linearly unstable but
{\it convectively} stable,
$Q_{\rm cong}  \in [Q_{\rm c3},Q_{\rm cv}]$, one has a spatial
{\it coexistence} HCT/OCT of states with HCT near the bottleneck and
OCT further upstream.
If, for yet higher flows, congested traffic is also convectively
unstable, 
the resulting oscillations lead to TSG or OCT.
In summary, the boundaries of the nonhysteretic transitions are given by
\begin{equation}
\begin{array}{ll}
\mbox{HCT}\leftrightarrow \mbox{HCT/OCT}:
       & \delta Q\approx Q_{\rm out}-Q_{\rm c3}, \\
\mbox{OCT} \leftrightarrow \mbox{HCT/OCT}: 
       & \delta Q\approx Q_{\rm out}-Q_{\rm cv}.
\end{array}
\end{equation}
Congested traffic of the HCT/OCT type is
frequently found in empirical data \cite{Kerner-wide}.
In the IDM, this frequent occurrence is reflected by the 
wide range of flows falling into this regime.
For the IDM parameters chosen here, we have
$Q_{\rm c3}=0$, and $Q_{\rm cv}=1050$ vehicles/h, i.e., {\it all}
congested states are linearly unstable and oscillations will develop
further upstream, while $Q_{\rm c3}$ is nonzero for the parameters of
Ref. \cite{TGF99-Treiber}.

Free traffic is (meta)stable in the overall system if it is 
(meta)stable in the bottleneck region. This means, a breakdown necessarily
takes place if the inflow $Q_{\rm in}$ exceeds the critical flow 
$[Q'_{\rm c2}(\delta Q) - Q_{\rm rmp}]$, where the linear stability threshold
$Q'_{\rm c2}(\delta Q)$ in the downstream region 
is some function of the bottleneck strength.
For the IDM with the parameters
chosen here, we have $Q'_{\rm c2}\approx Q'_{\rm out}$, {see}
Fig. \ref{fig:stab}(b). Then, the condition for the maximum inflow 
{allowing} for free traffic  simplifies to
\begin{equation}
\label{FT-CT}
\mbox{FT}\to\mbox{PLC or FT}\to\mbox{CT}: 
Q_{\rm in} \approx Q_{\rm out} - \delta Q,
\end{equation}
i.e., {it} is equivalent to relation (\ref{CT-PLC}).
In the phase diagram {of} Fig. \ref{phasediag}, 
this boundary is given by the dotted line.
For bottleneck strengths
$\delta Q \le 350$ vehicles/h, this line coincides with that of the
transition CT $\to$ PLC, in agrement with Eqs (\ref{FT-CT}) and
(\ref{CT-PLC}). For larger bottleneck {strengths,} the
approximation $\tilde{Q}_{\rm out}\approx Q'_{\rm out}$ used to derive
relation (\ref{CT-PLC}) is not fulfilled. 
%

\section{
\label{sec:emp}
Empirical Data of Congested Traffic States and their 
Microscopic Simulation}

We analyzed one-minute averages of
detector data from the German freeways
A5-South and A5-North near Frankfurt, A9-South
near Munich, and A8-East from Munich to Salzburg.
Traffic breakdowns occurred 
frequently on all four freeway sections.
The data suggest that the congested states
depend not only on the traffic situation but also
on the specific infra\-structure.

On the A5-North, we mostly found pinned localized clusters
(ten {times} during the observation period).
Besides, we observed 
moving localized clusters (two {times}), {triggered stop-and-go traffic}
(three {times}), and oscillating congested traffic (four {times}).

All eight recorded traffic breakdowns on the A9-South were to
oscillatory congested traffic, and all emerged upstream of 
intersections. The data of the A8 East showed OCT
with a more heavily congested HCT/OCT state propagating through it.
Besides this, we found  breakdowns to HCT/OCT
on the
A5-South (two {times}), one of them caused by lane closing due to
an external incident. 
In contrast, HCT states are often found on the Dutch freeway A9 
from Haarlem to Amsterdam behind an on-ramp with a very high inflow
\cite{Helb-book,TGF99-Treiber,GKT-scatter,Smulders1,Vladi-98}.
Before we present representative data for each traffic state,
some remarks about the presentation of the data are in order.

\subsection{Presentation of the Empirical Data}

In all cases, the traffic data were obtained
from several sets of 
double-induction-loop detectors recording, separately for each lane,
the passage times and velocities of all vehicles. 
Only aggregate information was stored.
On the freeways A8 and A9, the numbers of cars and trucks
that crossed a given detector on a given lane in each one-minute
interval,
and the corresponding average
velocities was recorded.
On the freeways A5-South and A5-North, the data
are available in form of a histogram for the velocity distribution.
Specifically, the measured velocities are divided into $n_r$ ranges
($n_r=15$ for cars and 12 for trucks),
and the number of cars and trucks driving in each range
are recorded for every minute.
This has the advantage that more ``microscopic'' information 
is given compared to
one-minute averages of the velocity.
In particular, the local traffic density 
$\rho^*(x,t)=Q/V^*$
could be estimated using the ``harmonic'' mean
$V^*= \erw{1/v_{\alpha}}^{-1}$ 
of the velocity, instead of the arithmetic mean
$\rho=Q/V$ with $V=\erw{v_{\alpha}}$.
Here, $Q$ is the traffic flow (number
of vehicles per time iterval), and $\erw{\cdots}$
denotes the {\it temporal} average  
over all vehicles $\alpha$ 
passing the detector within the given time interval.

{The harmonic mean value $V^*$ corrects for the fact
that the {\em spatial} velocity distribution differs from the {\em
locally measured} one} \cite{Helb-book}. {However,}
for better comparison with those freeway data, where this information
is not available, we
will use always the arithmetic velocity average $V$ in this paper.
Unfortunately, the velocity intervals of the A5 data 
are coarse. In particular, the 
lowest interval ranges from 0 to 20 km/h.
Because we used
the centers of the intervals as estimates for the velocity, there is
an artificial
cutoff in the corrsponding flow-density  diagrams \ref{HCTemptheo}(b),
and \ref{PLCMLCemp}(c).
Below the line $Q_{\rm min}(\rho)=V_{\rm min} \rho$ with
$V_{\rm min}= 10$ km/h.
Besides time series of flow and velocity
and flow-density diagrams, we present the data also
in form of three-dimensional plots of the
locally averaged velocity and traffic  density
 as
a function of position and time.
This representation is particularly useful to
distinguish the different congested states by their
qualitative spatio-temporal dynamics.
Two points are relevant, here.
First, the smallest time scale of the collective effects (i.e., the
smallest period of
density oscillations) is of the order of 3 minutes.
Second, the spatial resolution of the data is restricted to
typical distances between two neighboring detectors 
which are of the order of 1 kilometer.
To smooth out the  small-timescale fluctuations, and
to obtain a continuous function
$Y_{\rm emp}(x,t)$ from the one-minute values 
$Y(x_i,t_j)$ of detector $i$ at time $t_j$ with
$Y=\rho$, $V$, or $Q$, we applied
for all three-dimensional plots of {an} empirical quantity $Y$
the following smoothing and interpolation
procedure:
\begin{equation}
\label{smooth}
Y _{\rm emp}(x,t) = \frac{1}{N} \sum_{x_i} \sum_{t_j} Y(x_i,t_j)
\exp\left \{- \ \frac{(x-x_i)^2}{2\sigma_x^2}
            - \ \frac{(t-t_j)^2}{2\sigma_t^2} 
    \right\}.
\end{equation}
The quantity
\begin{equation}
\label{normalization}
N = \sum_{x_i} \sum_{t_j}
\exp\left \{ - \ \frac{(x-x_i)^2}{2\sigma_x^2} 
             - \ \frac{(t-t_j)^2}{2\sigma_t^2} 
    \right\},
\end{equation}
is a normalization factor.
We used smoothing times and length scales of
$\sigma_t=1.0$ min and
$\sigma_x=0.2$ km, {respectively}.
For {consistency,} we applied this smoothing operation 
also to the simulation results.
Unless explicitely stated otherwise, we will 
understand all empirical data as lane averages.

\subsection{\label{sec:HCT}
Homogeneous Congested Traffic}

Figure \ref{HCTemp} shows data of
a traffic breakdown on the A5-South on
August 6, 1998. 
Sketch  \ref{HCTemp}(a) shows the considered section. 
The flow data at cross section D11 {in} Fig. \ref{HCTemp}(b)
{illustrate} that, between 16:20 h and 17:30 h,
the traffic flow on the right lane
dropped to nearly zero.
For a short time interval between 17:15 h and 17:25 h 
also the flow on the
middle lane dropped to nearly zero.
Simultaneously, there is a sharp drop of the velocity at this cross
section on all lanes, cf. Fig. \ref{HCTemptheo}(d).
In contast, the velocities at the downstram
cross section D12 remained relatively
high  during the same time.
This suggests
a closing of the right lane at a location
somewhere between the detectors
D11 and D12. 

Figures \ref{HCTemptheo}(d) and (f) show that,
in most parts
of the congested region, there were little variations of the velocity.
The traffic flow remained relatively high, which is a signature of
synchronized traffic \cite{Kerner-sync}. 
In the immediate upstream (D11) and downstream (D12) neighbourhood
of the bottleneck, the amplitude of the fluctuations 
of traffic flow was
low as well, in particular, it was lower than in free traffic 
(time series at D11 and D12 for $t<$ 16:20 h, or $t>$ 18:00 h).
Further upstream in the congested region (D10), however,
the {fluctuation} amplitude increases.
After the bottleneck {was} removed at about 17:35 h, the 
previously fixed downstream front started moving
in the {\it up}stream direction at a characteristic velocity
of about 15 km/h \cite{Kerner-rehb98}.
Simultaneously, the flow
increased to about 1600 vehicles/h,
see plots \ref{HCTemptheo}(e) and \ref{HCTemptheo}(g).
After the congestion dissolved at about 17:50 h, the flow
dropped to about 900 vehicles/h/lane, which was the inflow at that time.

Figure \ref{HCTemptheo}(a) shows the flow-density 
diagram of the lane-averaged one-minute data.
In agreement with the absence of large oscillations ({like} stop-and-go
traffic), the regions of data points of free and congested traffic were 
clearly separated.
Furthermore, 
the transition from the free to the congested state and the reverse
transition showed a clear hysteresis.

The spatio-temporal 
plot of the local velocity in Fig. \ref{HCT3d}(a)
shows that 
the incident induced a breakdown to an extended state
of essentially homogeneous congested
traffic. Only near the upstream boundary, there were {small}
oscillations.
While the upstream front 
(where vehicles {entered} the congested region) propagated upstream,
the downstream front (where vehicles {could} accelerate into free traffic)
remained fixed at the bottleneck at $x\approx 478$ km.
In the spatio-temporal plot of the traffic flow Fig. \ref{HCT3d}(b),
one clearly can see the flow peak in the region $x>476$ km after the
bottleneck was removed.

We estimate now the point in the phase diagram to which 
this situation {belongs}. The average inflow $Q_{\rm in}$ ranging from
1100 vehicles/h at $t=$ 16:00 h to about 900 vehicles/h ($t=$ 18:00 h)
can be determined
from an upstream cross section which is not reached by the congestion,
in our case D6. Because the congestion
emits no stop-and go waves, we conclude that the free traffic
in the inflow region is stable,
$Q_{\rm in}(t)<Q_{\rm c1}$.
We estimate the bottleneck strength
$\delta Q = Q_{\rm out}-\tilde{Q}_{\rm out} \approx$ 700 vehicles/h
by identifying the time- and lane-averaged flow 
at D11 during the time of the incident 
(about 900 vehicles/h)
with the outflow $\tilde{Q}_{\rm out}$ from {the} bottleneck, and
the average flow of 1600 vehicles/h during the flow peak 
({when} the congestion dissolved) with
the {(universal) dynamical capacity $Q_{\rm out}$ 
on the homogeneous freeway (in the absence of a bottleneck-producing
incident).} For the short time interval where
two lanes were closed, we even have
$\tilde{Q}_{\rm out} \approx 500$ vehicles/h corresponding to
$\delta Q \approx 1100$ vehicles/h.
(Notice, that the lane averages {were always carried out 
over all} three lanes, also if lanes {were} closed.)
Finally, we conclude from the  oscillations near the upstream boundary of 
the congestion, that the congested traffic flow
$Q_{\rm cong}=(Q_{\rm out}-\delta Q)$ is linearly unstable, but
convectively stable.
Thus, the breakdown corresponds to the HCT/OCT regime.


{We simulated the situation with the IDM parameters from Table
\ref{tab:param}, with upstream boundary conditions
taken from the data of cross section D6
[cf. Fig. \ref{HCTemptheo} (h) and (i)] and homogeneous von Neumann
downstream boundary conditions.}
We implemented the temporary bottleneck by locally increasing the
model parameter $T$ to some value $T'>T$
in an 1 km long section centered around the 
location of the incident. 
This section represents the actually closed road section
and the merging regions upstream and downstream from it.
During the incident, we chose $T'$ such that the outflow 
$\tilde{Q}'_{\rm out}$ from the bottleneck agrees roughly with the
data of cross-section D11. 
At the beginning of the
simulated incident, we increased $T'$ abruptly from $T=1.6$ s to
$T'=5 s$, and decreased it linearly to 2.8 s during the time interval
(70 minutes) of
the incident. Afterwards, we assumed again $T'=T=1.6$ s.

The grey lines of Figs. \ref{HCTemptheo} (c) to \ref{HCTemptheo} (j)
show time series of the simulated velocity and flow at some
detector positions.
Figures \ref{HCT3d}(c) and (d) show plots of the smoothed
spatio-temporal velocity and flow, respectively.

Although, in the microscopic picture, the modelled increase of the
safe time headway is quite different from lane changes before a
bottleneck,
the qualitative dynamics is essentially
the same as that of the data. In particular,
(i) the breakdown occured immediately after the bottleneck has been
introduced. 
(ii) As long as the bottleneck was active,
the downstream front of the congested state
remained stationary and fixed at the bottleneck, while
the upstream front propagated further upstream.
(iii) Most of the congested region consisted of HCT,
but oscillations appeared near the upstream front. 
The typical period of the simulated oscillations
($\approx 3$ min), however,
{was} shorter than that of the measured data ($\approx $ 8 min).
(iv) As soon as the bottleneck was removed, the downstream front
propagated upstream with the well-known characteristic velocity
$v_g=15$ km/h,
and there was a flow peak in the downstream
regions until the congestion {had} dissolved, cf. Figs.
\ref{HCTemptheo}(c), \ref{HCTemptheo}(e), and \ref{HCT3d}(d).
During this time interval,
the velocity increased {\it gradually} to the value for free traffic.

{Some remarks on the apparently non-identical upstram boundary
conditions in the empirical and simulated plots \ref{HCT3d}(a) and
\ref{HCT3d}(c) are in order.
In the simulation,
the {\em velocity} relaxes quickly from its prescribed value at the upstream
boundary to
a value corresponding to {\it free} equilibrium traffic at the given 
inflow. This is a rather generic effect which also
occurs in macroscopic models \cite{numerics}.
The relaxation takes places within the boundary region 
$3 \sigma_x=0.6$ km needed 
for the smoothing procedure (\ref{smooth}) and is, therefore, 
not visible in the figures. Consequently, the boundary conditions for
the velocity look different, although they have been taken from the data.
In contrast, the {\em traffic flow} cannot
relax because of the conservation of the number of vehicles
\cite{numerics}, 
and the boundary conditions in the corresponding 
empirical and simulated plots look, therefore, consistent 
[see. Figs.~\ref{HCT3d} (b) and (d)].
These remarks apply also to all other simulations below.
}
\subsection{Oscillating Congested Traffic}

We now present data from a section of the A9-South near Munich.
There are two major intersections I1 and I2 with other 
freeways, cf. Fig. \ref{OCTemp}(a). In addition, the number of
lanes is reduced from three to two downstream of I2.
There are three further
small junctions between I1 and I2 which did not appear to be
dynamically relevant.
The intersections, however, were major bottleneck inhomogeneities.
Virtually on each weekday, traffic broke down  to 
oscillatory congested traffic
upstream of intersection I2. In addition, we recorded two breakdowns to OCT 
{upstream of} I1 during the observation period of 14 days.

Figure \ref{OCTemp}(b) shows a spatio-temporal 
plot of the smoothed velocity 
of the OCT state occurring upstream of I2 during the morning rush hour of 
October 29, 1998.
The oscillations with a period of about 12 min
are clearly visible in both the time series of the velocity
data, plots \ref{OCTemp}(d)-(f), and the flow,
\ref{OCTemp}(g)-(i). In contrast to the observations of
Ref. \cite{Kerner-wide}, the {density waves apparently} did not merge.
Furthermore, the velocity in the OCT state rarely exceeded 50 km/h, 
i.e., there was no free traffic between the clusters,
a signature of OCT in comparison with triggered stop-and go waves.
The clusters propagated upstream 
at a remarkably constant velocity of 15 km/h, which is 
nearly the same propagation velocity as that of the 
detached downstream front of the HCT state described above.

Figure \ref{OCTemp}(c) shows the 
flow-density diagram of this congested state. In contrast to the
diagram \ref{HCTemptheo}(b) for the HCT state, there is no
separation between the regions of free and congested
traffic. Investigating flow-density diagrams of many other occurrences 
of HCT and OCT, it turned out that this difference can also be used to
empirically distinguish HCT from OCT states.


Now we show that this breakdown to OCT can be qualitatively reproduced 
by a microsimulation with the IDM.
As in the previous simulation, we used empirical data for the upstream 
boundary conditions. {(Again, the velocity relaxes quickly to a local
equilibrium, and only for this reason it looks different from the
data.)} We implemented the 
bottleneck by locally increasing the safe time headway in the 
downstream region.
In contrast to the previous simulations, the 
local defect causing the breakdown was a permanent
inhomogeneity of the infrastructure (namely an intersection
and a reduction from three to two lanes) rather than 
a temporary incident. Therefore, we did not assume any time dependence
of the bottleneck.
As upstream boundary conditions, we chose the data of D20, 
the only cross section where
there was free traffic during the whole time interval considered here.
{Furthermore, we used homogeneous von Neumann boundary conditions at
the downstream boundary.}
Without assuming a higher-than-observed level of inflow, the
simulations showed no traffic breakdowns at all. 
Obviously, on the freeway A9 the capacity per lane is lower than
on the freeway A5 (which is several hundret kilometers apart). 
This lower capacity has been taken into account by a site-specific,
increased value of $T=2.2$\ s in the upstream region $x<-0.2$ km.
An even higher value of $T'=2.5$ s was used in the bottleneck region 
$x>0.2$ km, with a linear increase in the 400 m long transition zone.
The corresponding microsimulation is shown in Fig.~\ref{OCTtheo}.

In this way,
we obtained a qualitative agreement with  the A9 data.
In particular, 
(i) traffic broke down at the bottleneck spontaneously, in contrast to
the situation on the A5.
(ii) Similar to the situation on the A5,
the downstream front of the resulting OCT state was fixed at the
bottleneck while the upstream front propagated further upstream.
(iii) The oscillations showed no mergers and propagated with about 15
km/h in upstream direction. Furthermore, their 
period (8-10 min) is comparable 
with that of the data, and the velocity in the OCT region was
always much lower than that 
of free traffic.
(iv) After about 1.5 h, the upstream front reversed
its propagation direction and eventually dissolved.
The downstream front remained always fixed at the permanent inhomogeneity.
{Since, at no time, there is a clear transition from congested to free traffic
in the region upstream of the bottleneck (from which 
one could determine
$Q_{\rm out}$ and compare it with the outflow
$\tilde{Q}_{\rm out}\approx Q'_{\rm out}$ from the bottleneck), 
an estimate of the empirical bottleneck strength 
$(Q_{\rm out} - Q'_{\rm out})$ is
difficult. Only at D26, for times around 10:00 h, 
there is a region where the vehicles accelerate.
Using the corresponding traffic flow
as coarse estimate for $Q_{\rm out}$, and the minimum smoothed flow
at D26 (occurring between $t\approx$ 8:00 h and 8:30 h) as an estimate for 
$Q'_{\rm out}$,
leads to an empirical bottleneck strength
$\delta Q$ of 400 vehicles/h. This is consistent with
the OCT regime in the phase diagram Fig. \ref{phasediag}.
However, estimating the theoretical bottleneck strength directly from 
the difference $(T-T')$, using Fig. \ref{fig:Qout}, 
would lead to a smaller value.
To obtain a full quantitative agreement, it would probably be
necessary to calibrate more than just {\em one} IDM parameter to the
site-specific driver-vehicle behavior, or to
explicitely model the bottleneck by on- and off-ramps.}

\subsection{Oscillating Congested Traffic Coexisting with Jammed Traffic}

Figure \ref{flugzeug} shows an example of a more complex
traffic breakdown that occurred on the
freeway A8 East from Munich to Salzburg during the evening rush hour
{on} November 2, 1998. 
Two different kinds of bottlenecks were involved, (i) a relatively
steep uphill gradient from $x=38$ km to $x=40$ km
(``Irschenberg''), and (ii) an
incident leading to the closing of one of the three lanes between
the cross sections D23 and D24 from $t=$\,17:40 h until $t=$\,18:10 h.
The incident was deduced from the velocity and flow data of the
cross sections D23 and D24 as described in Section IV B.
As further inhomogeneity, there is a small junction at about
$x=41.0$ km. {However,} since the involved {ramp} flows were very small,
we {assumed}
that the junction {had no} dynamical effect.

The OCT state caused by the uphill gradient had the same qualitative
properties as that on the A9 South. In particular, the 
breakdown was triggered by a short flow peak corresponding 
 to a velocity dip in Plot
\ref{flugzeug}(b), the downstream
front was stationary, while the upstream front moved, 
and all oscillations propagated upstream with a constant velocity.
The combined HCT/OCT state caused by the incident had similar
properties {as} that on the A5 South. In particular, there was HCT
near the location of the incident,
corresponding to  the downstream boundary of 
the velocity plot \ref{flugzeug}(b),
while oscillations developped further upstream. Furthermore, similarly 
to the incident on the A5, the
downstream front propagated upstream as soon as the incident was 
{cleared.} 
The plot clearly shows, that the HCT/OCT state propagated 
seemingly unperturbed through the
OCT state upstream of the permanent uphill bottleneck. The 
upstream propagation velocity
$v_g=15$ km/h of {\it all} perturbations in the complex state was 
remarkably constant,
in particular that of (i) the upstream and downstream fronts
separating the
HCT/OCT state from free traffic (for $x>40$ km at $t\approx$\,17:40 h and
18:10 h, respectively), (ii) the fronts separating the HCT/OCT 
from the OCT state (35 km $\le x \le$ 40 km), and (iii) 
the oscillations within both the HCT and HCT/OCT states.
In contrast, the propagation velocity and direction of the
front separating the OCT from {\it free} traffic
varied with the inflow.

We simulated this scenario using empirical (lane averaged) data both
for the upstream and downstream boundaries.
For the downstream boundary, we used only the velocity information.
Specifically, 
when, at {some} time $t$,  a simulated vehicle $\alpha$ crosses 
the downstream boundary of the simulated section $x\in[0,L]$, 
we set its
velocity to that of the data, $v_{\alpha}=V_{D15}(t)$, if 
$x_{\alpha}\ge L$, and use the velocity and positional information 
of this vehicle to determine the 
acceleration of the vehicle $\alpha+1$ behind. Vehicle $\alpha$ is
taken out of the simulation as soon as vehicle
$\alpha+1$ has crossed the boundary. Then, the velocity of vehicle
$(\alpha+1)$ is set to the actual boundary value, and so on.
The downstream boundary conditions are only relevant for
the time interval around $t=18:00$ h where traffic near this boundary
is congested. For other time periods,
the simulation result is equivalent to using homogeneous
Von-Neumann downstream boundary conditions.

We modelled the stationary uphill bottleneck in the usual way by
increasing the parameter $T$ to a constant value $T'>T$ in the downstream 
region. The incident {was already reflected} 
by the downstream boundary conditions.
Figure \ref{flugzeug}(c) shows the simulation result in form
of a spatio-temporal plot of the smoothed velocity.
Notice that, by using only the
boundary conditions and a stationary bottleneck as 
specific information, we obtained a qualitative agreement 
of nearly all dynamical collective
aspects of the {whole complex scenario} described above.
In particular, for all times $t<$\,17:50 h, there was free traffic at both
upstream and downstream detector positions. Therefore, the boundary 
conditions (the detector data)
did not contain any explicit information about the
breakdown {to OCT inside the road section,} 
which nevertheless was reproduced correctly {as an emergent phenomenon}.


\subsection{Pinned and Moving Localized Clusters} 
%
Finally, we consider a 30 km long section
of the A5-North depicted in Fig. \ref{PLCMLCemp}(a).
On this section, we found  
one or more traffic breakdowns {on} six out of 21 days, all of them
Thursdays or Fridays. 
On three out of 20 days,
we observed one or more 
stop-and-go waves separated by free traffic. 
The stop-and go waves were
triggered near an intersection and agreed qualitatively with the
TSG state of the phase diagram. 
{On} one day, two isolated {density} clusters
propagated through the considered region
and did not trigger any secondary clusters, which is consistent
with moving localized clusters (MLC) and will be discussed below.
Moving localized clusters were observed quite
frequently on this freeway section \cite{Kerner-rehb96}.
Again, they have a constant upstream propagation speed of about 15 km/h,
and a characteristic outflow \cite{KK-94}.
In addition, we found four breakdowns to OCT, and ten occurrences of
pinned localized clusters (PLC).
The PLC states emerged either
at the intersection I1 (Nordwestkreuz Frankfurt), or
1.5 km downstream of intersection I2 (Bad Homburg) at cross section 
D13. 
Furthermore, the downstream fronts of all four OCT states
were fixed at the latter location. 

{On August 6, 1998, we found an interesting transition from an OCT state
whose downsteam front was at D13, to a TSG state with a downstram
front at intersection I2 (D15).
Consequently, we conclude
that, around detector D13, there is a stationary
flow-conserving bottleneck with a stronger effect than the
intersection itself.
Indeed, there is an uphill section and a relatively sharp 
curve at this location of the A5-North, which may be the reason for the
bottleneck.
The sudden change of the active bottleneck on August 6 can be explained
by perturbations and the hysteresis associated with breakdowns.
}

{The different types of traffic breakdowns are
consistent with the relative locations 
of the traffic states in the
$(Q_{\rm in},\delta Q)$ space of the phase
diagram in Fig. \ref{phasediag}.
Three of the four occurrences of OCT and two of the three TSG states
were on Fridays (August 14 and  August 21, 1998), on which traffic flows
were about 5\%
{\it higher} than on our reference day (Friday, August 7, 1998), which
will be discussed in detail below. 
Apart from the coexistent PLCs and MLCs
observed on the reference day, all PLC states occurred on Thursdays,
where average traffic flows were about 5\%
{\it lower} than on the reference day.
No traffic breakdowns were observed on Saturdays to Wednesdays, 
where the traffic flows were at least 10\% 
lower compared to the reference day.
As will be shown below, for complex bottlenecks like intersections,
the coexistence of MLCs and PLCs
is only possible for flows just above those triggering pure
PLCs, but below those triggering OCT states.
So, we have, with increasing flows, the sequence FT, PLC, MLC-PLC, and 
OCT or TSG states, in agreement with the theory.
}

Now, we discuss the traffic breakdowns in August 7, 1998 in detail.
Figure \ref{PLCMLCemp}(b) shows the situation 
from $t=$ 13:20 h until 17:00 h in form of a spatio-temporal
plot of the smoothed density. 
During the whole time interval, there was a pinned localized cluster 
at cross section D13. 
Before $t=$\,14:00 h, the PLC state
showed distinct oscillations (OPLC), while it was essentially
stationary (HPLC) afterwards. 
Furthermore, two moving localized clusters (MLC) of unknown origin 
propagated through nearly the whole displayed section 
and also through a 10 km long downstream section (not shown
here) giving a total of at least 30 km.
Remarkably, as they crossed the PLC at D13,
neither of the congested states  seemed to be affected.
This complements the observations of Ref. \cite{Kerner-rehb96},
desribing MLC states that propagated unaffected through 
intersections {in the absence of} PLCs.
{As soon as the first MLC state reached
the location of the on-ramp of intersection I1 ($x=488.8$ km, 
$t\approx$ 15:10 h), it triggered
an additional pinned localized cluster, which dissolved at
$t\approx$ 16:00 h. The second MLC dissolved as soon as it reached
the on-ramp of I1 at $t\approx 16:40$ h.}

Figure \ref{PLCMLCemp}(c) demonstrates that
the MLC and PLC states have characteristic signatures also in the
empirical flow-density diagram.
As is the case for HCT and OCT, the PLC state is characterized
by a two-dimensional flow-density regime (grey
squares). In contrast to the former states, however, there is no flow
reduction (capacity drop) with respect to 
free traffic (black bullets). As is the case for flow-density diagrams
of HCT {compared to} OCT, it is expected that HPLCs are
characterized by an isolated region, while the points
of OPLC lie in a region which is connected to the region
for free traffic.
During the periods were the MLCs crossed the PLC at D13,
the high traffic flow of the PLC state dropped drastically,
and the traffic flow had essentially the property of the MLC,
{see} also the velocity plot \ref{PLCMLCemp}(e).
Therefore, we omitted in the PLC data the 
points corresponding to these intervals.

The black bullets for densities $\rho>30$ vehicles/km
indicate the region of the MLC (or TSG) states.
Due to the aforementioned difficulties in determining the traffic
density for very low velocities,
the theoretical line $J$ given by 
$Q_{\rm J}(\rho)= Q_{\rm out} 
 - (Q_{\rm out}-Q_{\rm jam})(\rho-\rho_{\rm out})/
                          (\rho_{\rm jam}-\rho_{\rm out})$
(see Ref.~\cite{Kerner-wide} and Fig.~\ref{PLCMLCemp}(b))
is hard to {\em find empirically.} In any case, {the data suggest that 
the line $J$ would lie} below the PLC region.


To simulate this scenario it is important that the PLC 
states occurred in or near the freeway intersections.
Because at both intersections, the off-ramp is upstream of the on-ramp 
[Fig. \ref{PLCMLCemp}(a)], the local flow at these locations is lower.
In the following, we will investigate the region around I2.
{During the considered time interval, the average traffic flow
of both, the on-ramp and the off-ramp was about 300 vehicles per hour
and lane. With the exception of the time intervals, during which
the two MLCs pass by, we have about
1200 vehicles per hour and lane at I2 (D15), 
and 1500 vehicles per hour and lane upstram (D16) and
downstream (D13) of I2.
This corresponds to an {\it increase} of the effective 
capacity {by} $\delta Q\approx -300$ vehicles per hour and lane
in the region between the off-ramp and the subsequent on-ramp.}

In the simulation, we captured this 
qualitatively by {\it decreasing}
the parameter $T$ in a section 
$x\in[x_1,x_2]$ upstream of the empirically observed PLC
state. The hypothetical bottleneck located at D13, i.e., 
about 1 km upstream of the on-ramp, {was neglected}. 
{Using real traffic flows as upstream and downstram
boundary conditions and varying only the model parameter
$T$ within and outside of the intersection,
we could not obtain
satisfactory simulation results.
This is probably because of the relatively high and fluctuating 
traffic flow on this highway. It remains to be shown if 
simulations with other model parameters can successfully reproduce the 
empirical data when applying real boundary conditions. 
Now, we show that the
main {\it qualitative} feature on this highway, namely, the coexistence 
of pinned and moving localized clusters can, nevertheless,
be captured by our model.
For this purpose, we assume a constant inflow $Q_{\rm in}=1390$
vehicles per hour and lane
to the freeway, with the corresponding equilibrium velocity. 
We initialize the PLC by a triangular-shaped density peak in the
initial conditions,
and initialize the MLCs by reducing the velocity at the downstream
boundary to $V=12$ km/h during two five-minute intervals (see caption 
of Fig. \ref{PLCMLCmic}). 
}
%

Again, we obtained a qualitative agreement with the observed dynamics.
In particular, the simulation showed that also an 
{\it increase} of the local capacity
in a bounded region
can lead to pinned localized clusters.
Furthermore, the regions of the MLC and PLC
states in the flow-density diagram were reproduced
qualitatively, in particular, the coexistence of
pinned and moving localized clusters.
We did not observe such a coexistence
in the simpler system 
{underlying the phase diagram in Fig.~\ref{phasediag}, which 
did not include a second low-capacity stretch 
upstream of the high-capacity stretch}.

{To explain the coexistence of PLC and MLC in the more complex
system consisting of one high-capacity stretch in the middle of
two low-capacity stretches, it is useful to interpret the inhomogeneity not 
in terms of a
local capacity {\it increase} in the region $x\in[x_1,x_2]$, but 
as a capacity {\it decrease} for $x<x_1$ and $x>x_2$.
(For simplicity, we will not explicitely include the 400 m long
transition regions of capacity increase at $x_1$ and decrease at $x_2$
in the following discussion.) 
Then, the location $x=x_2$ can be considered as the beginning
of a bottleneck, as in the system underlying the phase
diagram.
If the width
$(x_2-x_1)$ of the region with locally increased capacity
is {\it larger than the width of PLCs}, such clusters are
possible under the same conditions as in the standard phase diagram.
In particular, traffic in the standard system is stable in regions upstream of
a PLC, which is the reason why any additional
MLC, triggered somewhere in the downstream region $x>x_2$
and propagating upstream, will vanish as soon as it crosses 
the PLC at $x=x_2$.
However, this disappearance is not instantaneous, but the MLC will continue to
propagate upstream for an additional 
flow-dependent ``dissipation distance'' or ``penetration depths''.
If the width $(x_2-x_1)$ is
{\it smaller than the dissipation distance for MLCs},
crossing MLCs
will not fully disappear before they reach the upstream 
region $x<x_1$. There,
traffic is metastable again, so that the MLCs can persist.
Since, in the metastable regime, the outflow of MLCs is equal to their
inflow (in this regime, MLCs are equivalent to
``narrow'' clusters, cf. Ref. \cite{KK-94}), the passage of the MLC
does not change the traffic flow at the position of the PLC,
which can, therefore, persist as well.

We performed several simulations varying the inflow within the range
where PLCs are possible.
For smaller inflows, the dissipation distance became smaller than
$(x_2-x_1)$, and the moving localized cluster was absorbed within
the inhomogeneity.
An example for this can be seen in Fig. \ref{PLCMLCemp}(b) at $t\approx$
16:40 h and $x\approx$ 489 km.
Larger inflows lead to an extended OCT state
upstream of the capacity-increasing defect, which is
also in accordance with the observations.
}

\section{Conclusion}

\begin{figure}
\begin{center}
\includegraphics[width=85mm]{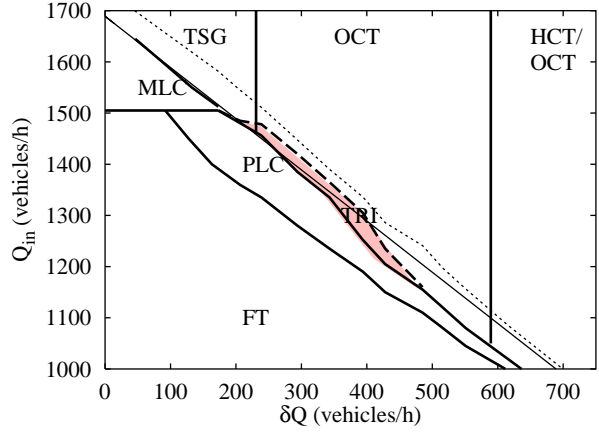}
\end{center}
\caption[]{
\label{phasediag}
Phase diagram resulting from IDM simulations of an open 
system with a flow-conserving bottleneck.
The bottleneck is realized by an increased IDM parameter $T$
in the downstream region, cf. Eq. (\protect\ref{Tx}).
The control parameters are the traffic flow $Q_{\rm in}$ and
the bottleneck strength $\delta Q$, see Eq. (\protect\ref{deltaQ}).
The solid thick lines separate the congested
traffic states TSG, OCT, HCT,
PLC, and MLC (cf. the main text and Fig. \protect\ref{3dphases}) 
and free traffic (FT)
as they appear after adiabatically increasing the inflow to the value
$Q_{\rm in}$ and applying a large perturbation afterwards
(history ``C'' in the main text).
Also shown is
the critical downstream flow $Q'_{\rm c2}(\delta Q)$ (thin solid line), 
below which free
traffic (FT) is (meta-)stable, and the maximum downstream flow
$Q'_{\rm max}(\delta Q)$ (thin dotted) below which (possibly unstable)
equilibrium traffic exists. 
Traffic is bistable for inflows above the lines FT-PLC or
FT-MLC (whichever is lower), and below $Q'_{\rm c2}(\delta Q)$.
In the smaller shaded region, traffic is tristable and
the possible states FT, PLC, or OCT depend on the previous history (see 
the main text). For history ``C'' we obtain OCT in this region.
}
\end{figure}

\begin{figure}
\begin{center}
\includegraphics[width=85mm]{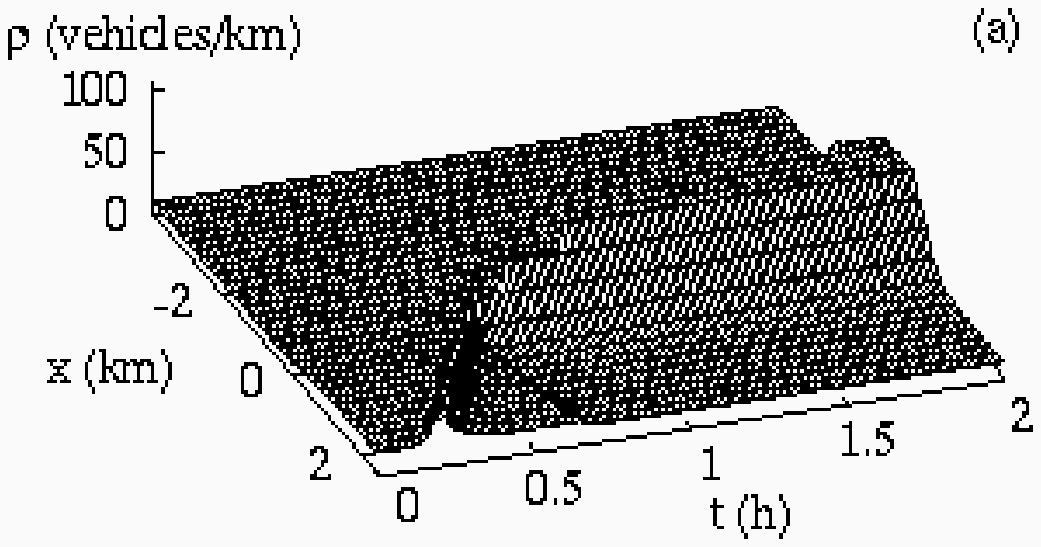}\\[0mm]
\includegraphics[width=85mm]{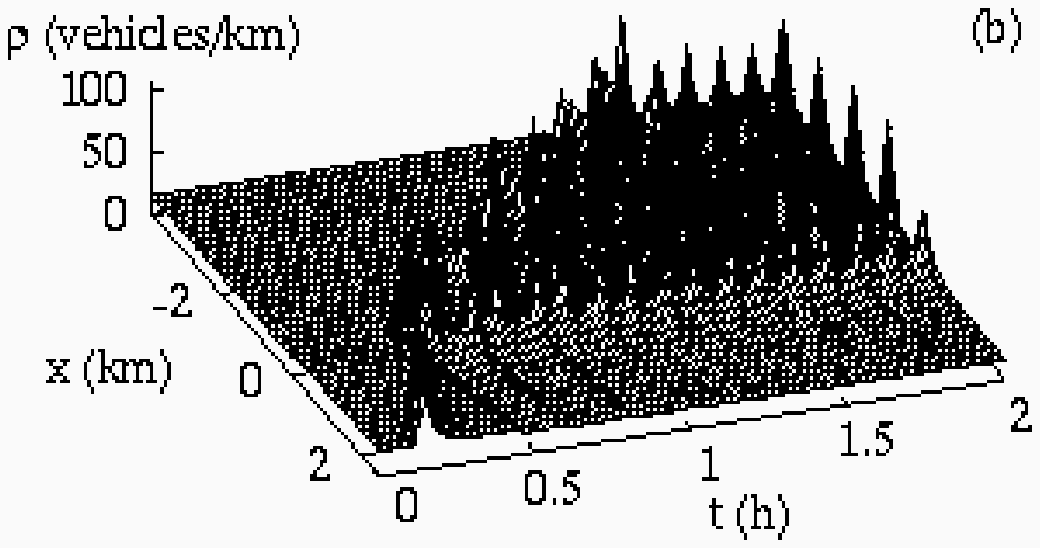}\\[0mm]
\includegraphics[width=85mm]{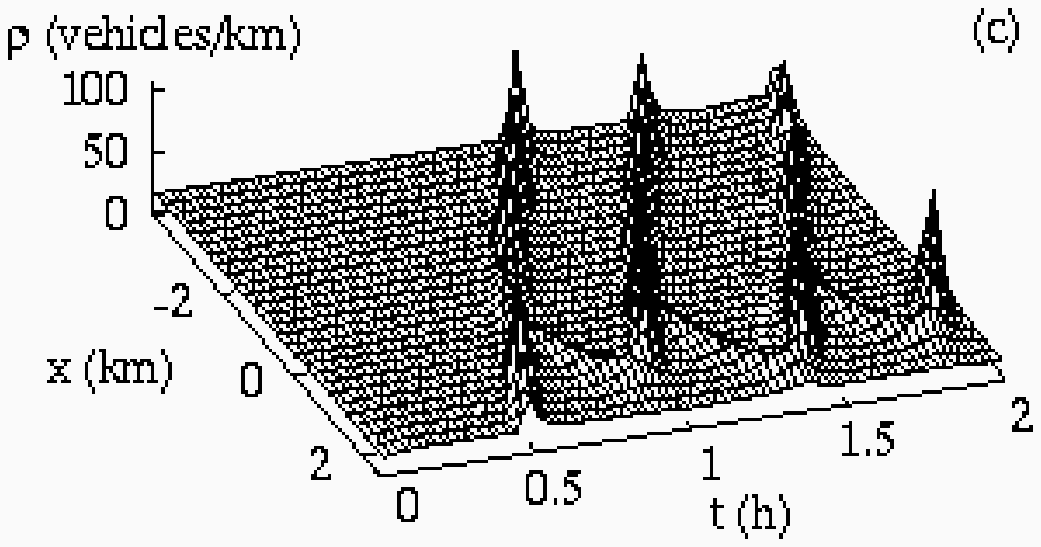}\\[0mm]
\includegraphics[width=85mm]{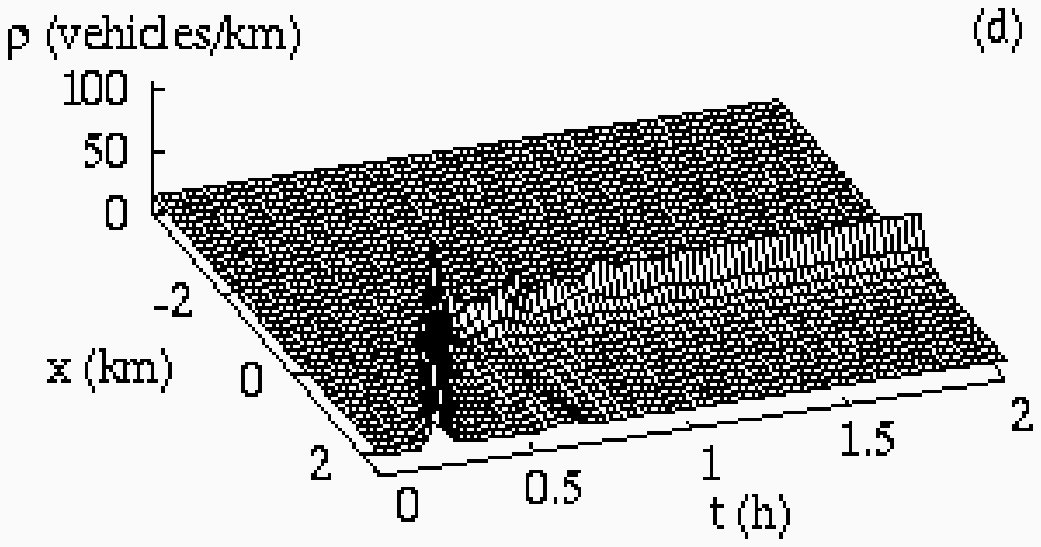}\\[0mm]
\includegraphics[width=85mm]{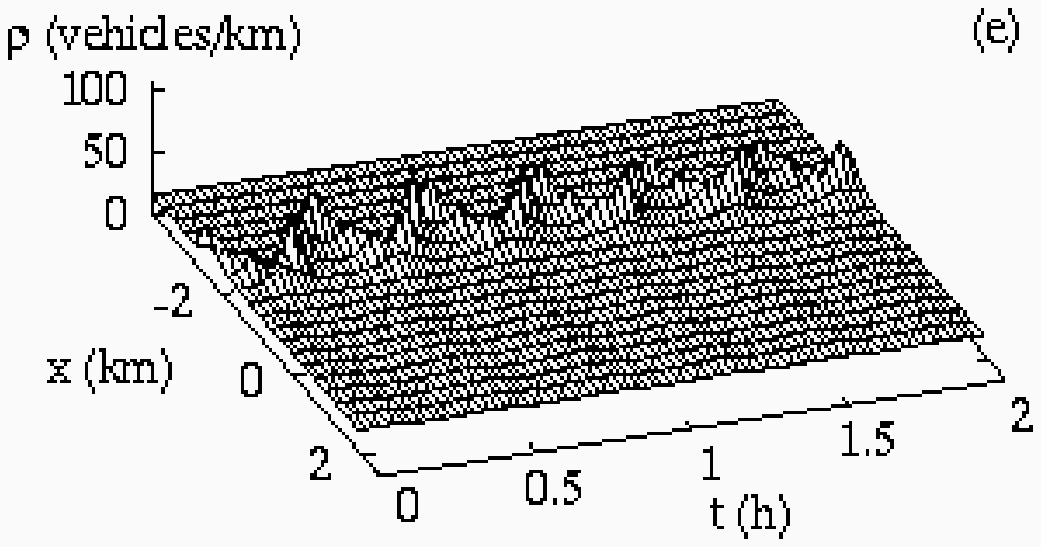}
\end{center}
\caption[]{
\label{3dphases}
Spatio-temporal density plots of the traffic states appearing in the
phase diagram of Fig.~\protect\ref{phasediag}.
(a) Homogeneous congested traffic (HCT), 
(b) oscillating congested traffic (OCT),
(c) triggered stop-and-go waves (TSG), 
(d) (stationary) pinned localized cluster (SPLC), and
(e) oscillatory pinned localized cluster (OPLC).
{The latter two states are summarized as pinned localized clusters 
(PLC).}
After a stationary state of free traffic has developed,
a {density} wave is introduced through the downstream
boundary ({or initial conditions})
which eventually triggers the breakdown 
(History ``C'', cf. the main text).
}
\end{figure}

\begin{figure}

\begin{center}
\includegraphics[width=85mm]{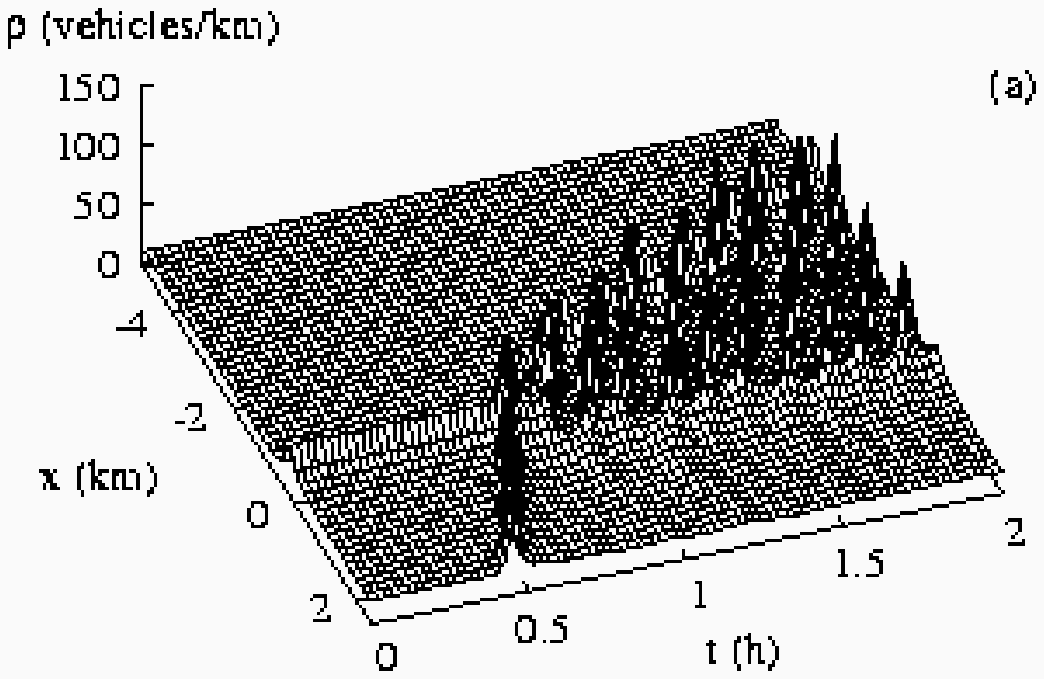}\\[0mm]
\includegraphics[width=85mm]{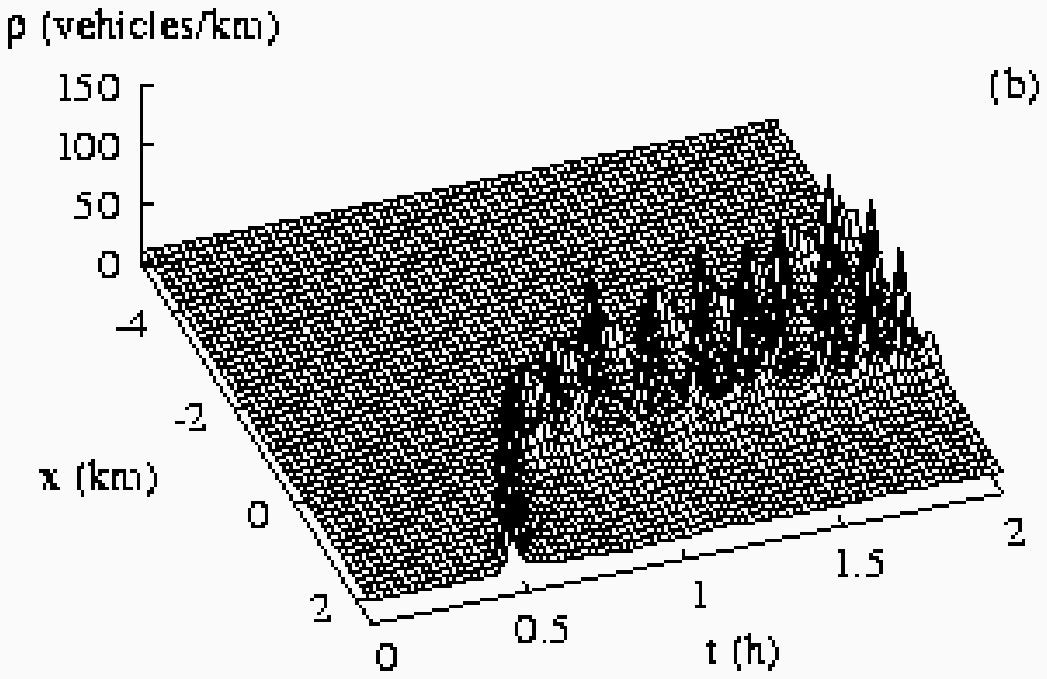}\\[0mm]
\includegraphics[width=85mm]{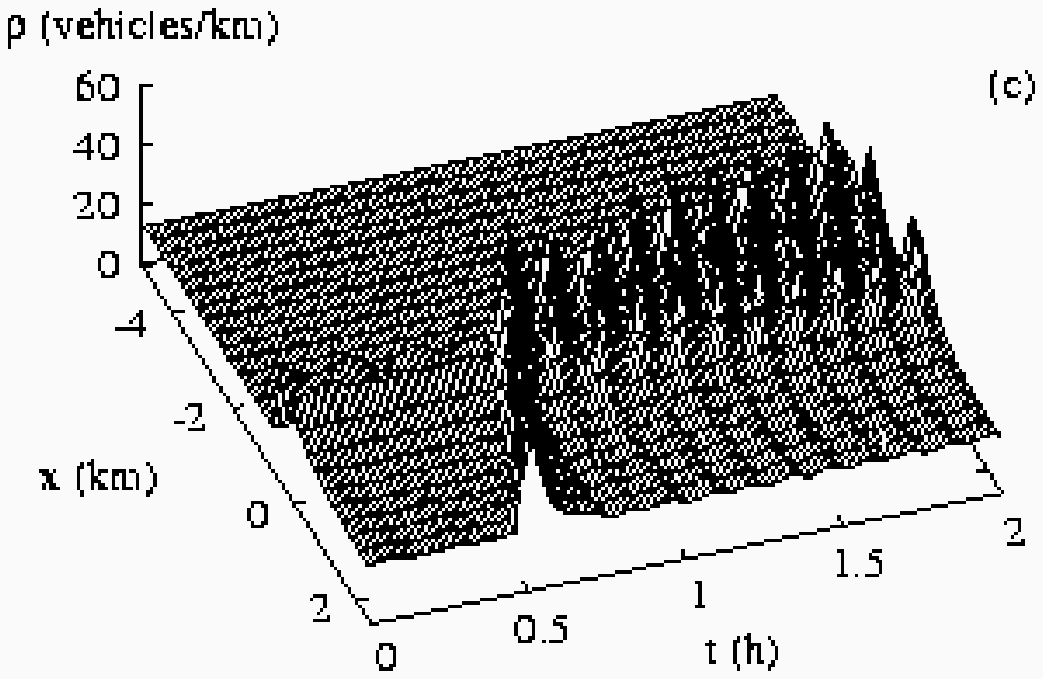}
\end{center}
\caption[]{
\label{tristab}
Spatio-temporal density plots {for the same phase point in}
the tristable traffic regime, {but different histories}.
(a) Metastable PLC and stable OCT.
The system is the same as in Fig. \protect\ref{3dphases} with
$Q_{\rm in}=1440$ vehicles/h and $T'=1.95$ s ($\delta Q=270$ 
vehicles/h). 
The {metastable} PLC is triggered by a triangular-shaped density peak 
in the initial conditions ({of} total width 600 m,
centered at $x=0$), {in which} the 
density rises from 14 vehicles/km to 45 vehicles km.
The OCT is triggered by a {density} wave introduced by 
the downstream boundary conditions.
(b) Same system as in (a), but {starting with} metastable free traffic. 
Here, {the transition to OCT is triggered by a density wave coming 
from the downstream boundary.}
(c) Similar behavior as in (a), but for the GKT model
(with model parameters $v_0=120$ km/h,
$T=1.8$ s, $\tau=50$ s, $\rho_{\rm max}=130$ vehicles/km,
$\gamma=1.2$, $A_0=0.008$, and
$\Delta A=0.008$, cf. Ref. \protect\cite{GKT}).
}
\end{figure}

\begin{figure}

\begin{center}

\includegraphics[width=85mm]
       {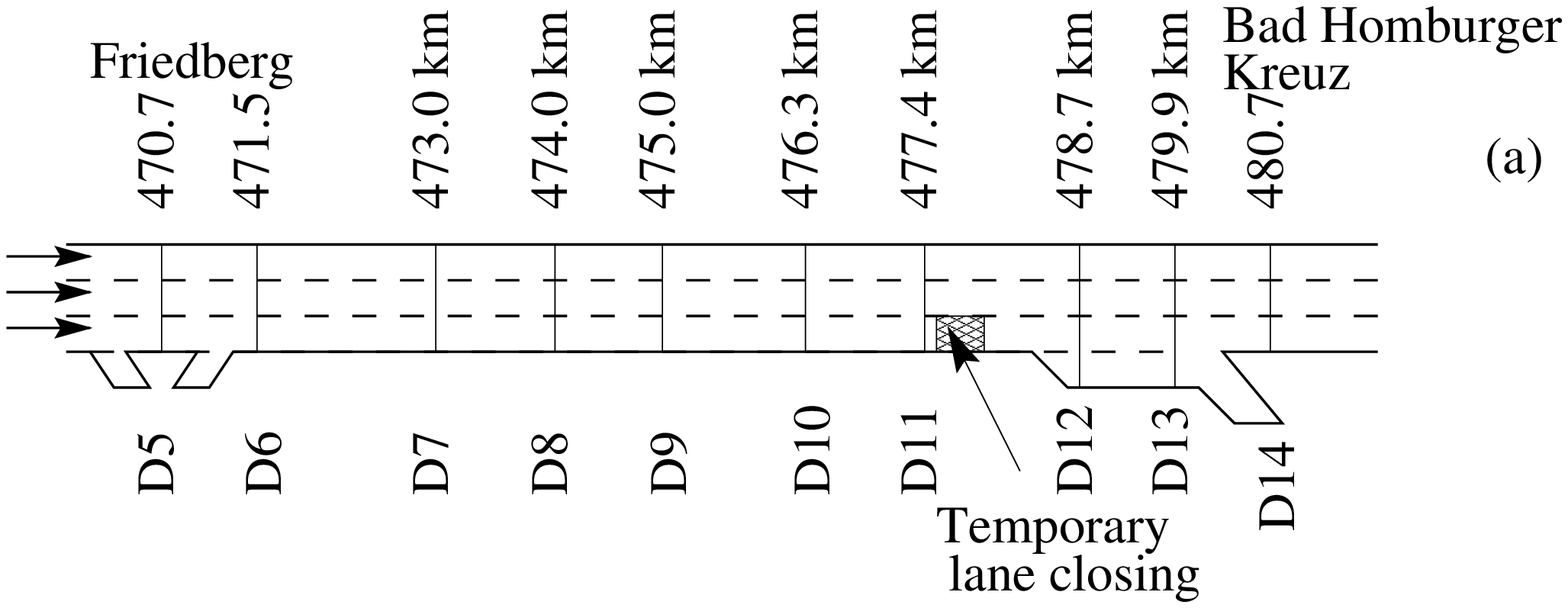} \\
\includegraphics[width=85mm]
       {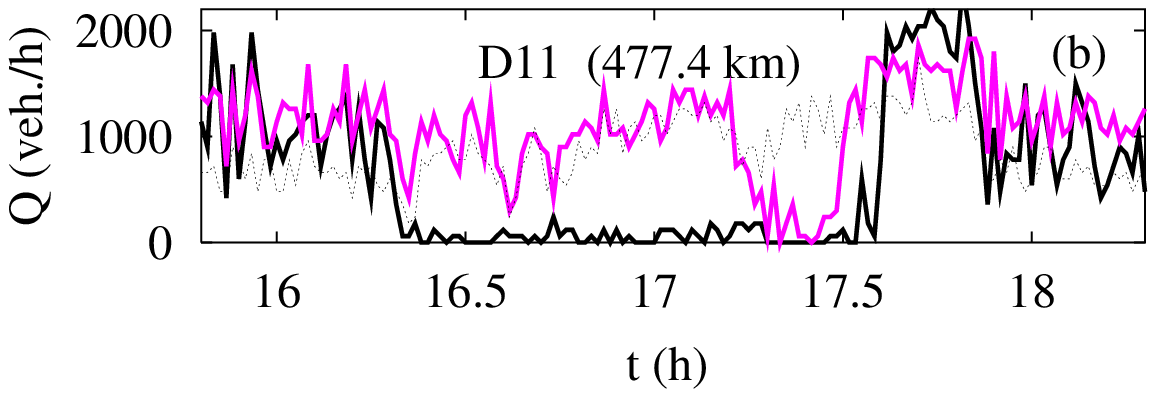}
\end{center}

\vspace*{3mm}

\caption[]{\label{HCTemp}
Traffic breakdown to nearly homogeneous congested
traffic on the freeway A5-South near Frankfurt
triggered by a temporary incident between 16:20 and 17:30
{on} Aug. 6, 1998 between the cross sections D11 and D12.
(a) Sketch of the freeway. (b) Flows at cross section D11
on the right lane 
(solid black), middle 
lane (grey), and left lane (dotted).
}
\end{figure}

\begin{figure}

\begin{center}
\includegraphics[width=160mm]{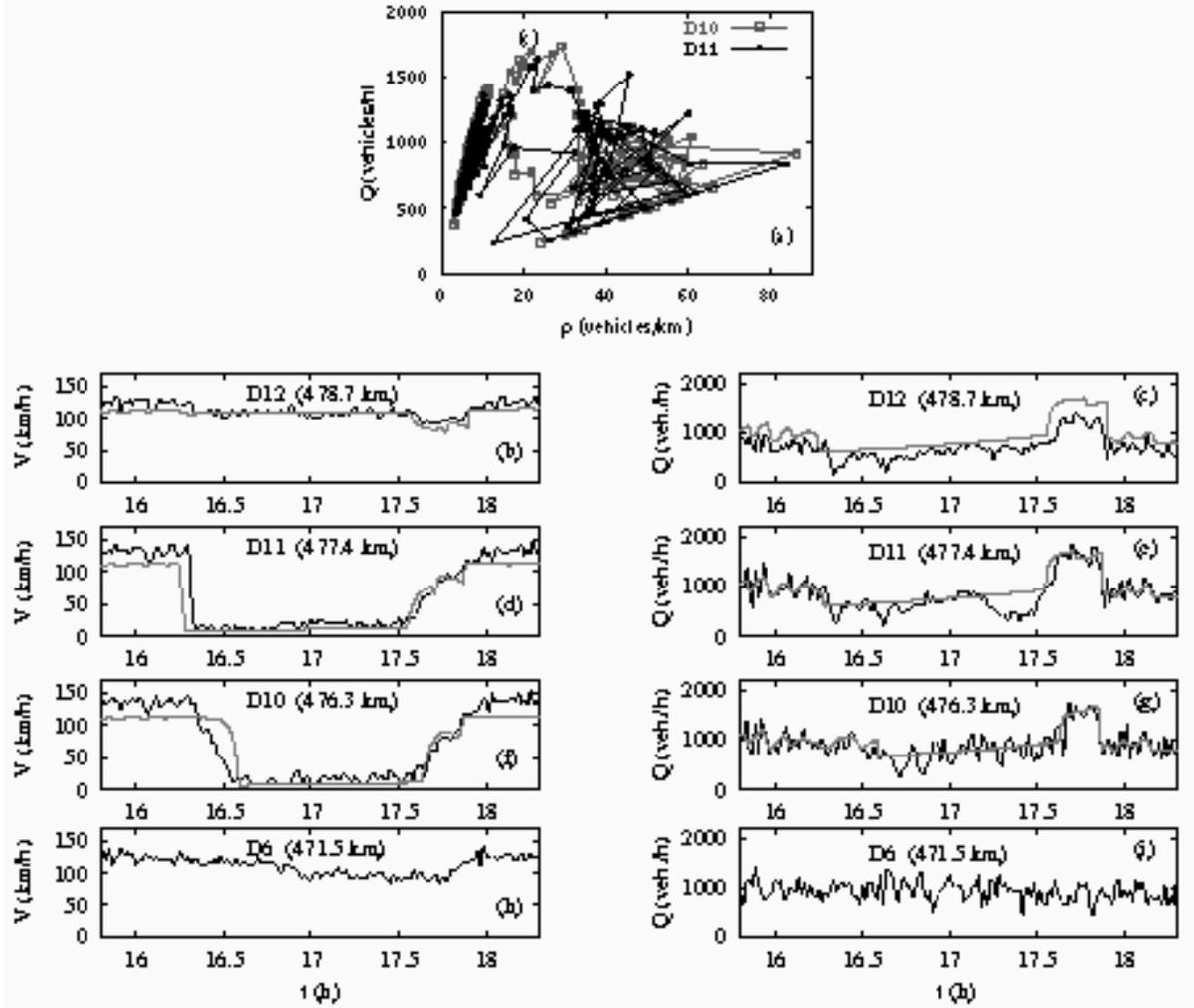} 
\end{center}

\vspace*{0mm}

\caption[]{\label{HCTemptheo}
{\bf [Please display in two columns as indicated]}
Details of the traffic breakdown
depicted in
Fig. \protect\ref{HCTemp}.
(a) Flow-density diagram of the traffic breakdown on the A5 South, 
(b)-(g) temporal evolution of velocity and flow
at three locations near the perturbation, and (h), (i) inflow boundary
conditions taken from cross section D6.
Besides the data (black lines), the results of the microsimulation are
shown (grey).
All empirical quantities are averages over one minute and over
{all} three lanes.
}
\end{figure}

\begin{figure}

\begin{center}
\includegraphics[width=160mm]{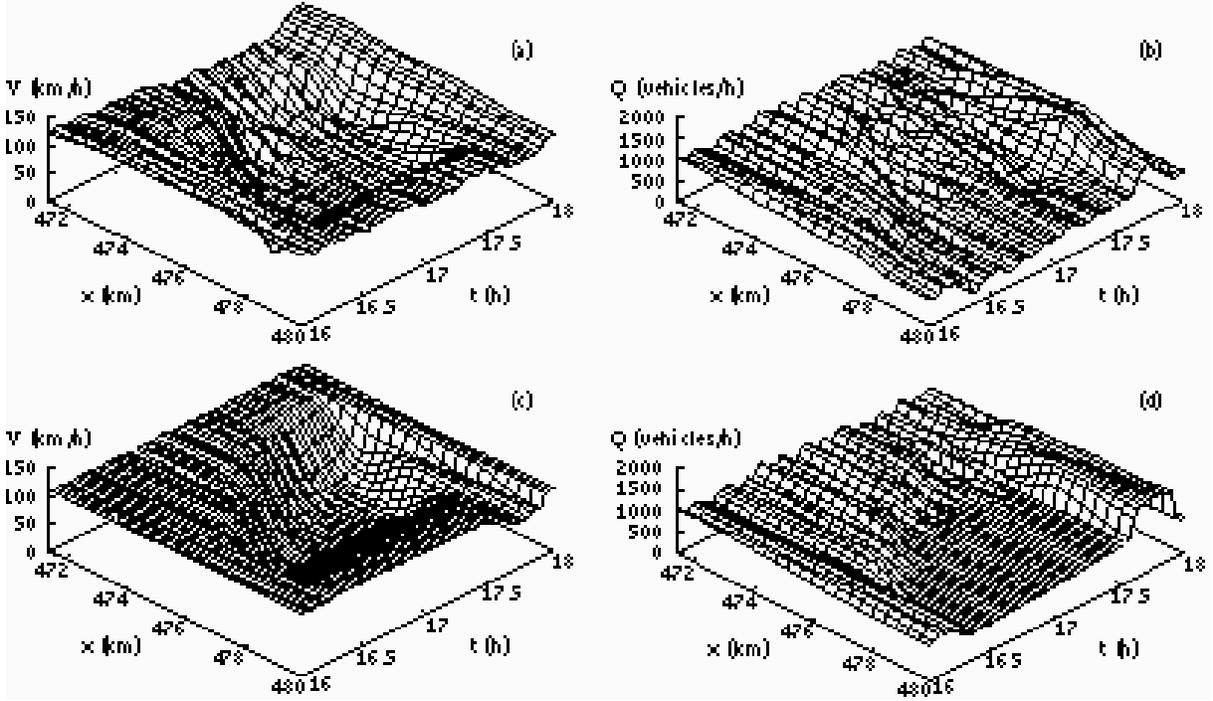} 

\end{center}

\caption{
\label{HCT3d}
{\bf [Please display in two columns as indicated]}
(a), (b) Smoothed spatio-temporal velocity 
and flow from the data
of the traffic breakdown depicted in
Figs.~\protect\ref{HCTemp} and \protect\ref{HCTemptheo}. 
(c), (d) Corresponding IDM microsimulation with the parameter set from
Table \ref{tab:param}.
The upstream boundary conditions for velocity and traffic flow were
taken from cross section D6. Because of the fast relaxation of the
velocity to the model's equilibrium value, 
the upstream boundary conditions for the empirical velocity
plots (a) seem to be
different from the simulation (c) (see main text).
Homogeneous von Neumann  boundary conditions
were assumed downstream.
The temporary lane closing is modelled by
locally increasing the IDM parameter $T$ 
in a 1000 m long section centered at $x=478$ km during the time
interval 16:20 h $\le t\le$ 17:30 h of the incident.
}
\end{figure}

\begin{figure}

\begin{center}
\vspace*{-5mm}
\includegraphics[width=160mm]{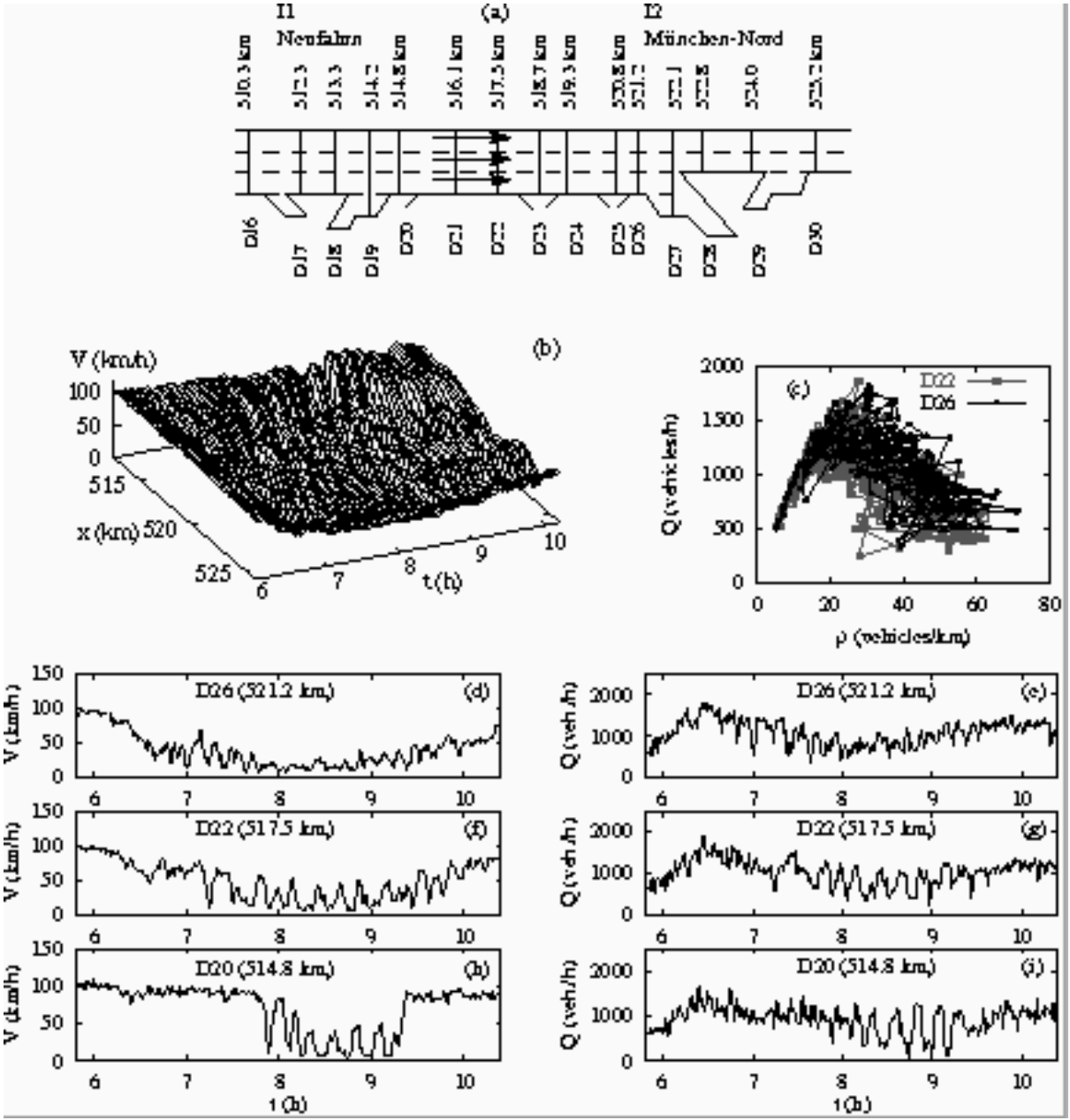} 
%
\end{center}

\caption[]{\label{OCTemp}
{\bf [Please display in two columns as indicated]}
Traffic breakdown to oscillating congested traffic in the evening rush 
hour of October 29, 1998 on the freeway A9-South near Munich.
(a) Sketch of the considered section with the cross sections D16 to
D30
and their positions in kilometers.
The small on- and off-ramps between
I1 and I2, which have been neglected in our simulation, are indicated
by diagonal lines.
(b) Spatio-temporal plot of the smoothed 
lane-averaged velocity. (c) Flow-density diagram obtained from
two detectors in
the congested region. 
(d)-(i) Lane-averaged 1-minute data of
velocities and flows.
}
\end{figure}

\begin{figure}

\begin{center}
\includegraphics[width=95mm]{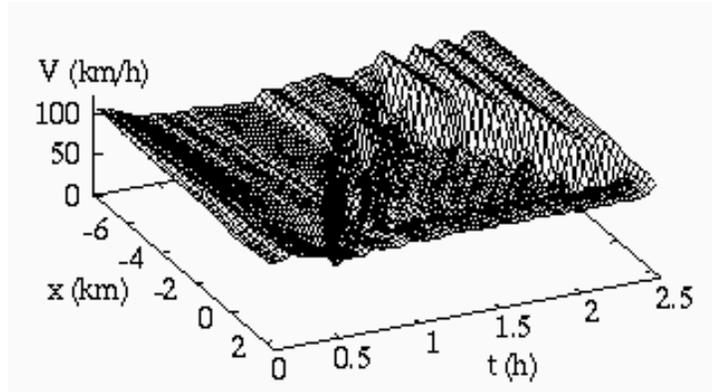} \\
\end{center}

\caption[]{\label{OCTtheo}
{Smoothed spatio-temporal velocity plot from a qualitative 
IDM microsimulation of the situation depicted in 
Fig. \protect\ref{OCTemp}. As inflow boundary conditions, we used the 
traffic flow data of
cross section D20. 
Homogeneous von Neumann
boundary conditions were assumed at the downstream boundary.
The inhomogeneity was implemented by a local, but time-independent
increase of $T$ 
in the region $x\ge 0$ km from
$T=2.2$ s to $T'=2.5$ s.}
}
\end{figure}

\begin{figure}

\begin{center}
\includegraphics[width=85mm]
      {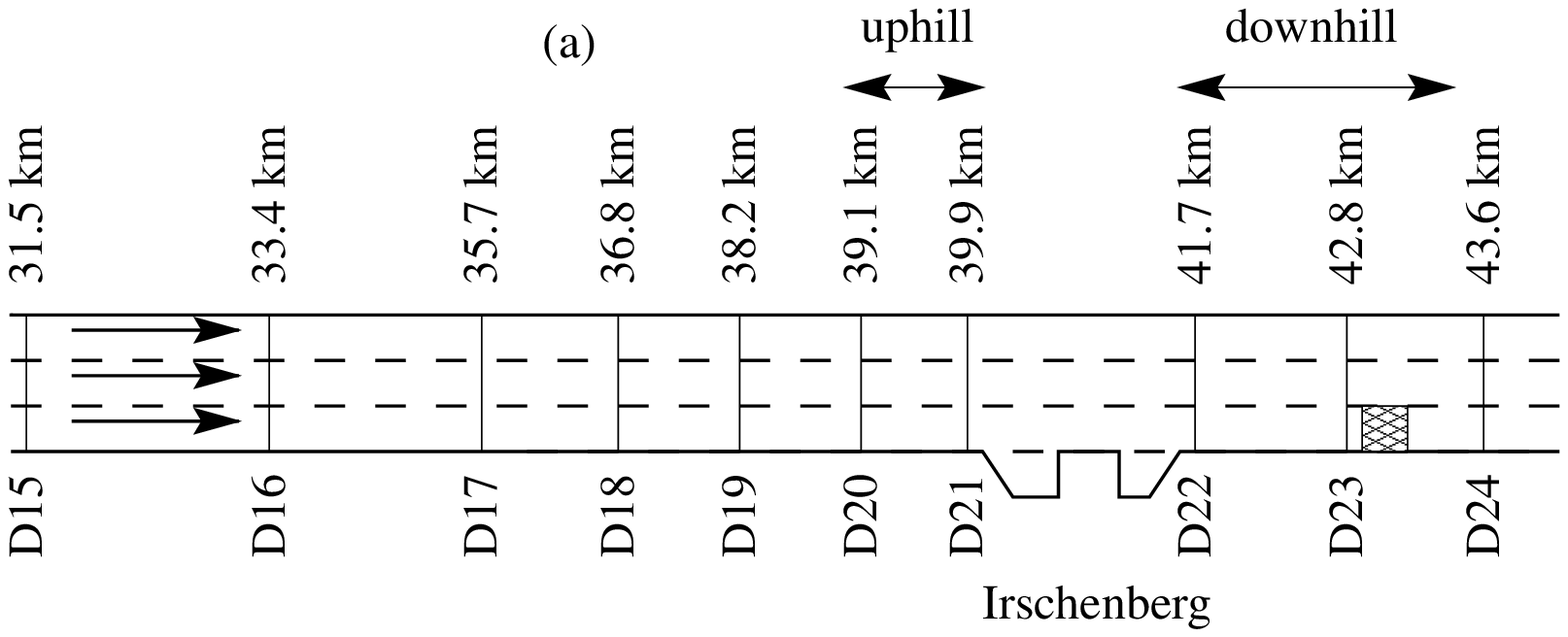} \\
\vspace*{-3mm}
\hspace*{0mm}
\includegraphics[width=90mm]{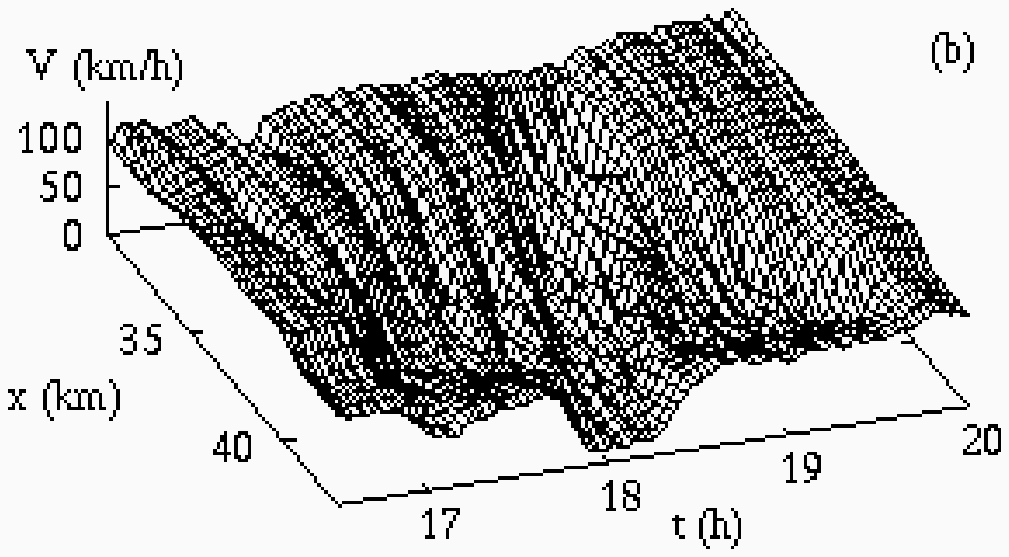}\\
\vspace*{0mm}
\includegraphics[width=90mm]{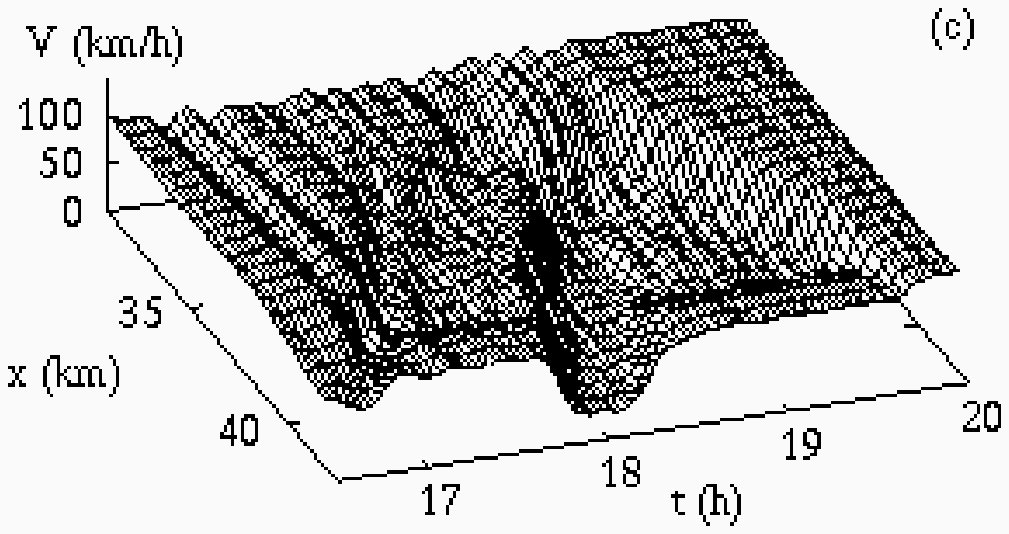}\\
\hspace*{-10mm}
\end{center}

\caption[]{\label{flugzeug}
Oscillating congested traffic (OCT) on an uphill
section of the freeway A8
East (near Munich). (a) Sketch of the section
with the cross sections D15 to D24 and
their positions in kilometers. (b) Smoothed
lane-averaged empirical velocity. An incident
leading to a temporary lane closing between D23 and D24
(near the downstream boundaries of the plots) induces even
denser congested traffic that propagates through the OCT region.
(c) Microsimulation using the data of cross sections D15 and D23 as
upstream and downstream boundary conditions, respectively.
The uphill section is modelled by linearly increasing the safe time headway
from $T=1.6$ s (for $x<39.3$ km) to $T'=1.9$ s  (for $x>40.0$ km).
{As in the previous microsimulations, the {\em velocity} near the upstram 
boundary relaxes quickly and,
therefore, seems to be inconsistent with the empirical values
(see main text).}
}
\end{figure}

In this paper, we investigated, {to what extent}
the phase diagram Fig,
\ref{phasediag} can serve as a general description of 
collective traffic dynamics in open, inhomogeneous systems.
The original phase diagram was formulated for on-ramps 
and resulted
from simulations with macroscopic models \cite{Phase,Lee}.
By simulations with a new car-following model we showed that
one can obtain the same phase diagram from microsimulations.
This includes even such subtle details as the small region of
tristability. 
The proposed intelligent-driver model (IDM) 
is simple, has only a few intuitive
parameters with realistic values,
reproduces a realistic collective dynamics,
and also leads to a plausible ``microscopic''
acceleration and deceleration behaviour of single drivers.
An interesting open question is whether the phase diagram
can be reproduced also with cellular automata.

We generalized the phase diagram from on-ramps
to {other kinds of} inhomogeneities. Microsimulations
of a flow-conserving
bottleneck  realized by
a locally increased safe time headway suggest that, with respect to
collective effects outside of the immediate neighbourhood of the
inhomogeneity, all types of bottlenecks  can be characterized by a single
parameter, the bottleneck strength. This means, that the type of
traffic breakdown depends {essentially} on 
the two control parameters
of the phase diagram {only,} namely the traffic flow, 
and the bottleneck strength. {However, in some multistable regions,
the history (i.e., the previous traffic dynamics) matters as well.}
We checked this also by macroscopic simulations with 
the same type of flow-conserving inhomogeneity and with
microsimulations using a locally decreased desired velocity as
bottleneck \cite{TGF99-Treiber}. In all cases, we
obtained qualitatively the same phase diagram.
What remains to be done is to confirm the phase diagram also
for microsimulations of on-ramps. These can be implemented
either by explicit multi-lane car-following
models \cite{Howe,Applet-engl},
or, in the framework of single-lane
models, by placing additional
vehicles in suitable gaps between vehicles in the ``ramp'' region.

By presenting empirical data of congested traffic, we showed that
all congested states proposed {by} the phase diagram were observed in
reality, among them localized and extended states which can be
stationary as well as oscillatory, furthermore, 
moving or pinned localized clusters (MLCs or PLCs, respectively). 
The data suggest that the typical {kind} of traffic
congestion depends on the specific freeway. This is in accordance
with other observations, for example, moving localized clusters on the
A5 North \cite{Kerner-rehb96}, or homogeneous congested traffic
(HCT) on the A5-South
\cite{Kerner-sync}. {In contrast to another empirical study
\cite{Kerner-wide}, the frequent oscillating states (OCT) 
in our empirical data
did not show mergings of density clusters, 
although these can be reproduced with our model with other
parameter values \cite{coexist}.}

The relative positions of the various observed
congested states in the phase diagram were
consistent with the theoretical predictions.
{In particular, when increasing the {\it traffic flow} on the freeway,
the phase diagram predicts (hysteretic) transitions from free
traffic to PLCs, and then
to extended congested states. 
By ordering the various forms of congestion on the A5-North with respect to
the average
traffic flow, the observations agree with these predictions. Moreover,
given an extended congested state
and increasing the {\it bottleneck strength}, the phase
diagram predicts
(non-hysteretic) transitions from triggered stop-and-go waves
(TSG) to OCT, and then to HCT.
To show the qualitative agreement with the data, we had to estimate 
the bottleneck strength $\delta Q$. This was done directly by identifying
the bottleneck strength with ramp flows, e.g.,
on the A5-North, 
or indirectly, by comparing
the outflows from congested traffic with and without a
bottleneck, e.g., for
the incident on the A5-South. With OCT and TSG on the A5-North, but
HCT on the A5-South, where the bottleneck strength was much higher,
we obtained again the right behavior.}
However, one needs a larger base of data to determine an empirical phase 
diagram, in particular with its boundaries between the different
traffic states. Such a phase diagram has been {proposed} 
for a Japanese highway \cite{Lee-emp}. 
Besides PLC states, many breakdowns on 
this freeway lead to extended congestions with fixed
{downstream {\it and} upstream} fronts. We did not observe such states 
on German freeways
and believe that the fixed upstream fronts were the result of a further
inhomogeneity, but this remains to be investigated.

\begin{figure}
\begin{center}
\includegraphics[width=85mm]
      {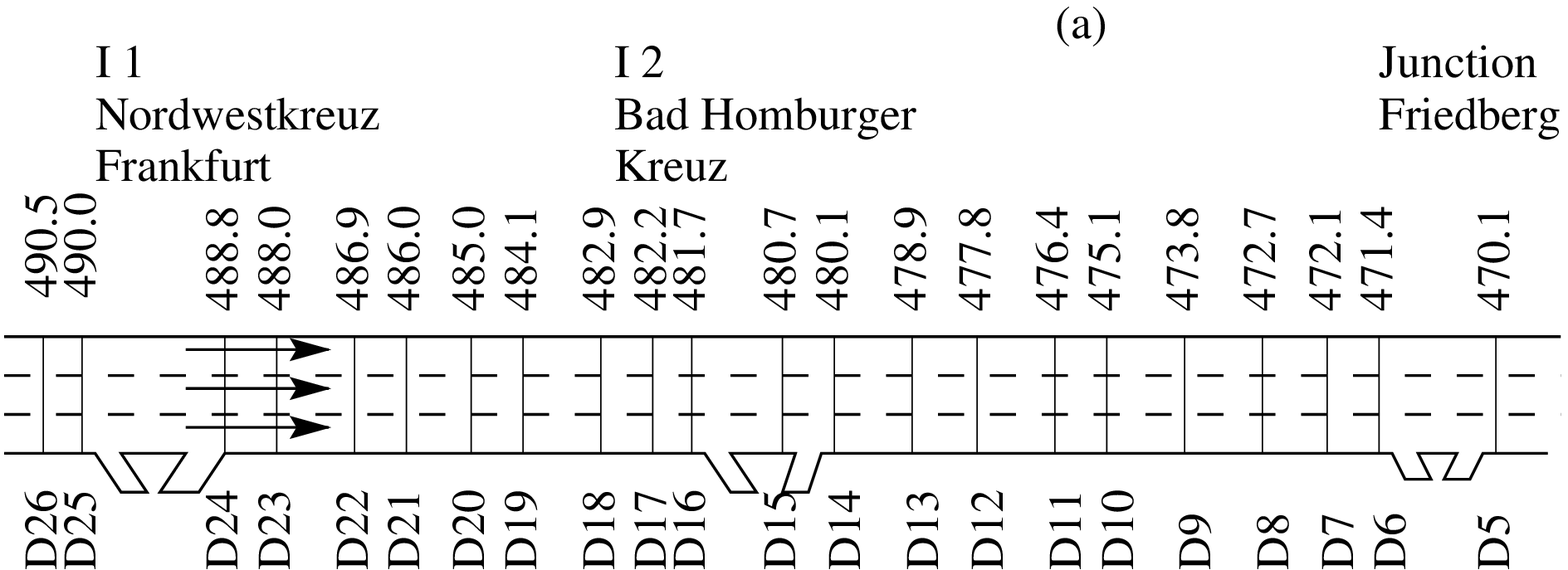}
\vspace*{0mm}\\

\hspace*{-5mm}
\includegraphics[width=95mm]{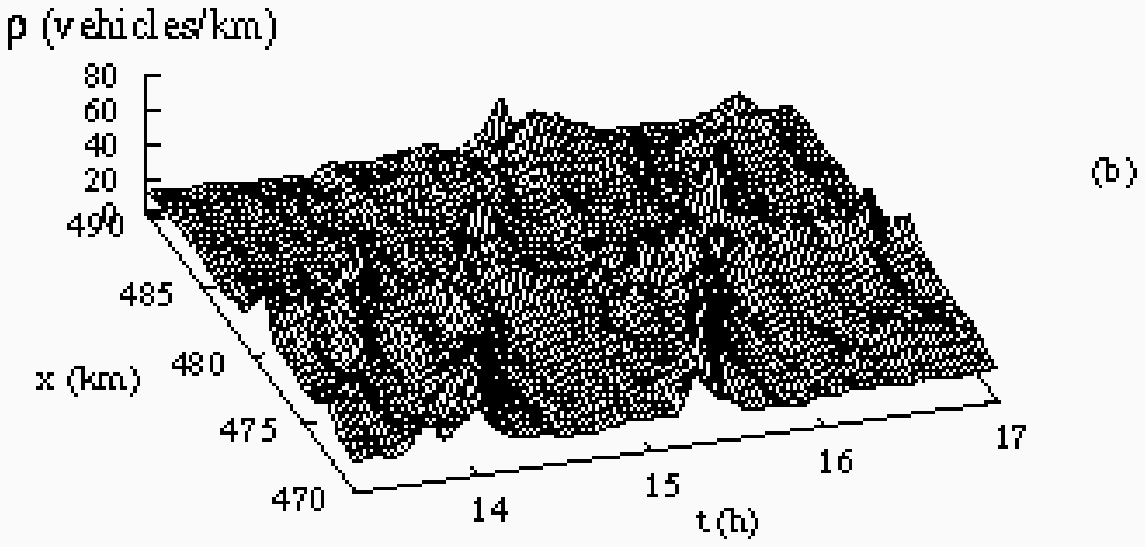} \\
\vspace*{-5mm}
\includegraphics[width=80mm]{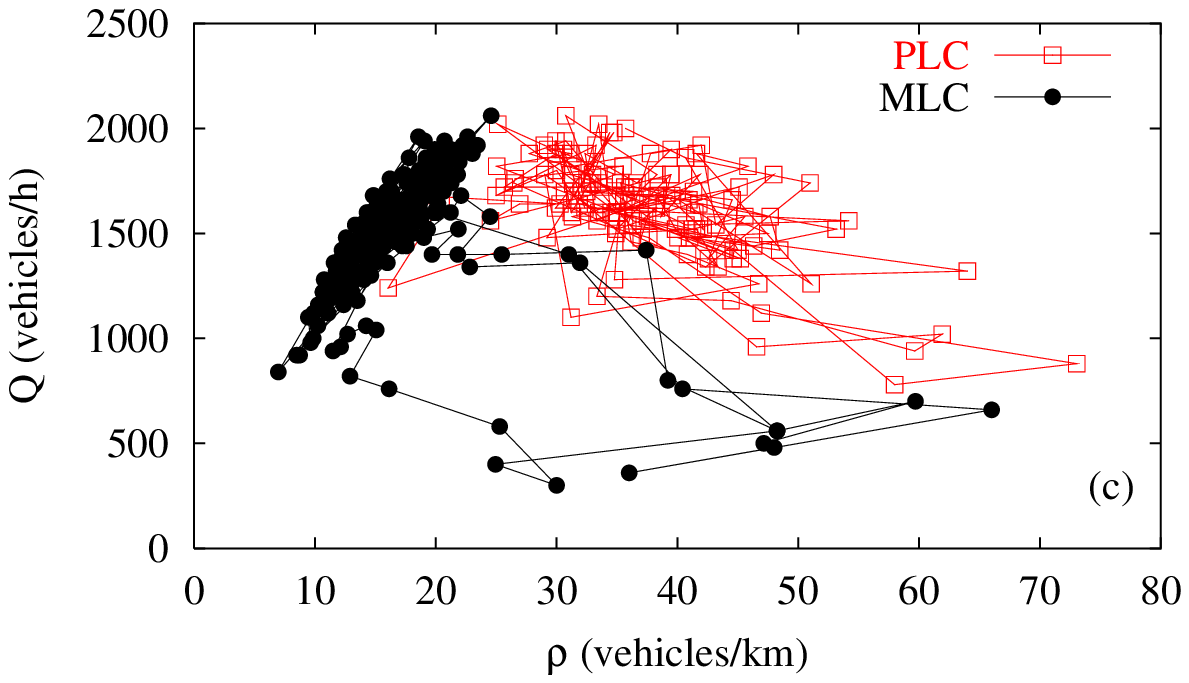} \\ 
\includegraphics[width=80mm]{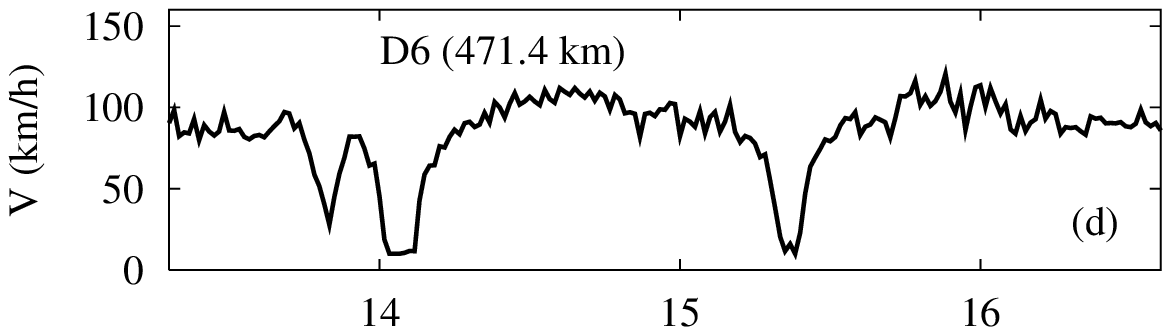}\\
\vspace*{-2mm}
\includegraphics[width=80mm]{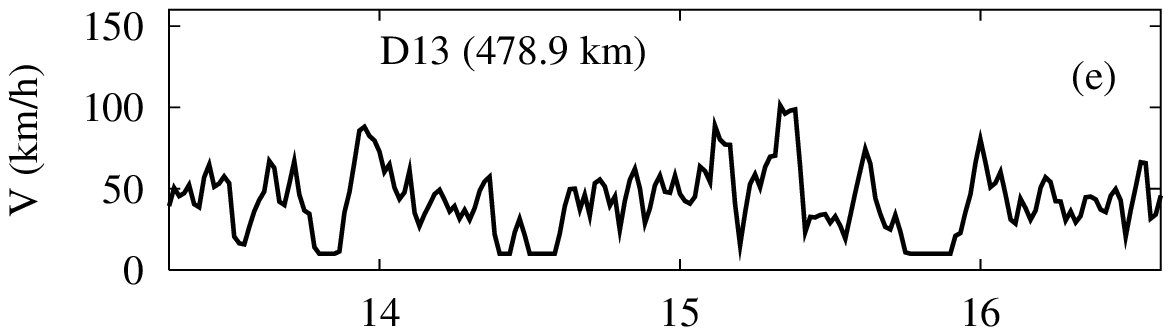}\\
\vspace*{-2mm}
\includegraphics[width=80mm]{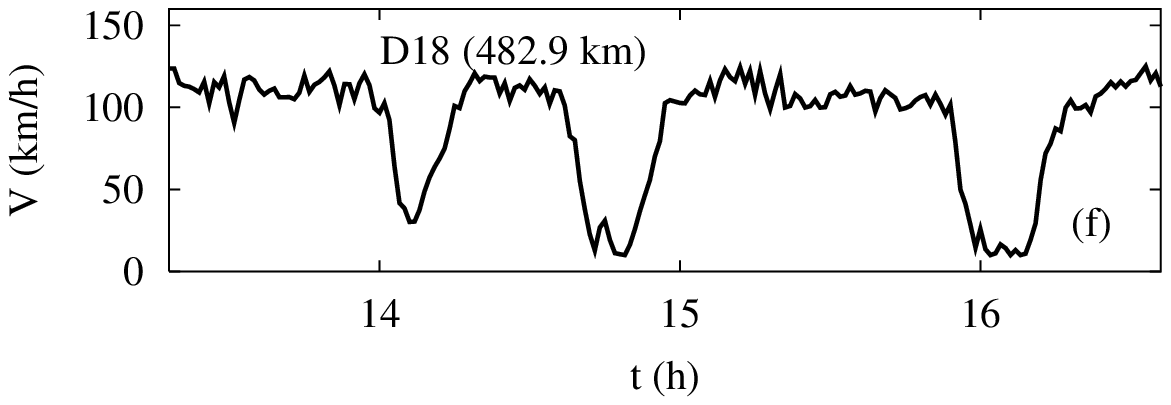}
\end{center}
\vspace*{0mm}

\caption{\label{PLCMLCemp}
Data of two moving localized
clusters (MLCs), and 
pinned localized clusters (PLC) on the freeway A5-North near Frankfurt.
(a) Sketch of the infrastructure with the positions of the cross
sections D5 - D16 in kilometers. (b) Spatio-temporal plot of the
density. (c)
Flow-density diagram {for} detector D6, where there are only stop-and-go 
waves ($\Box$), 
and {for detector} D13 ($\bullet$), {where we
have omitted data points during the time intervals when
the MLCs passed by.}
(d)-(f) Temporal evolution of the velocity (one-minute data)
at the location of the PLC at D13, and upstream (D18)
and downstream (D6) of it.
}
\end{figure}

\begin{figure}

\begin{center}
\includegraphics[width=105mm]{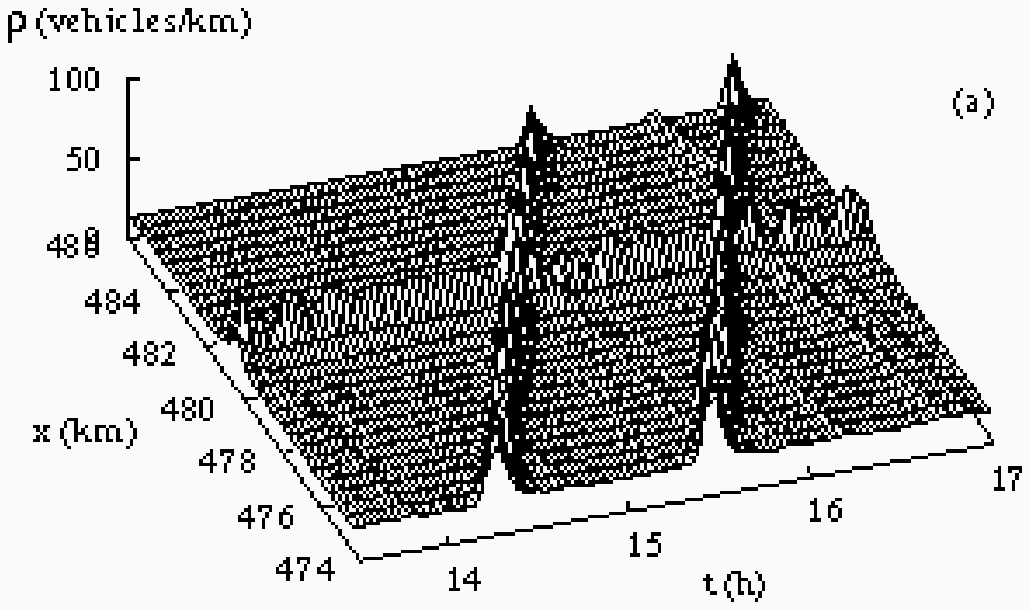}  \\[0mm]
\includegraphics[width=85mm]{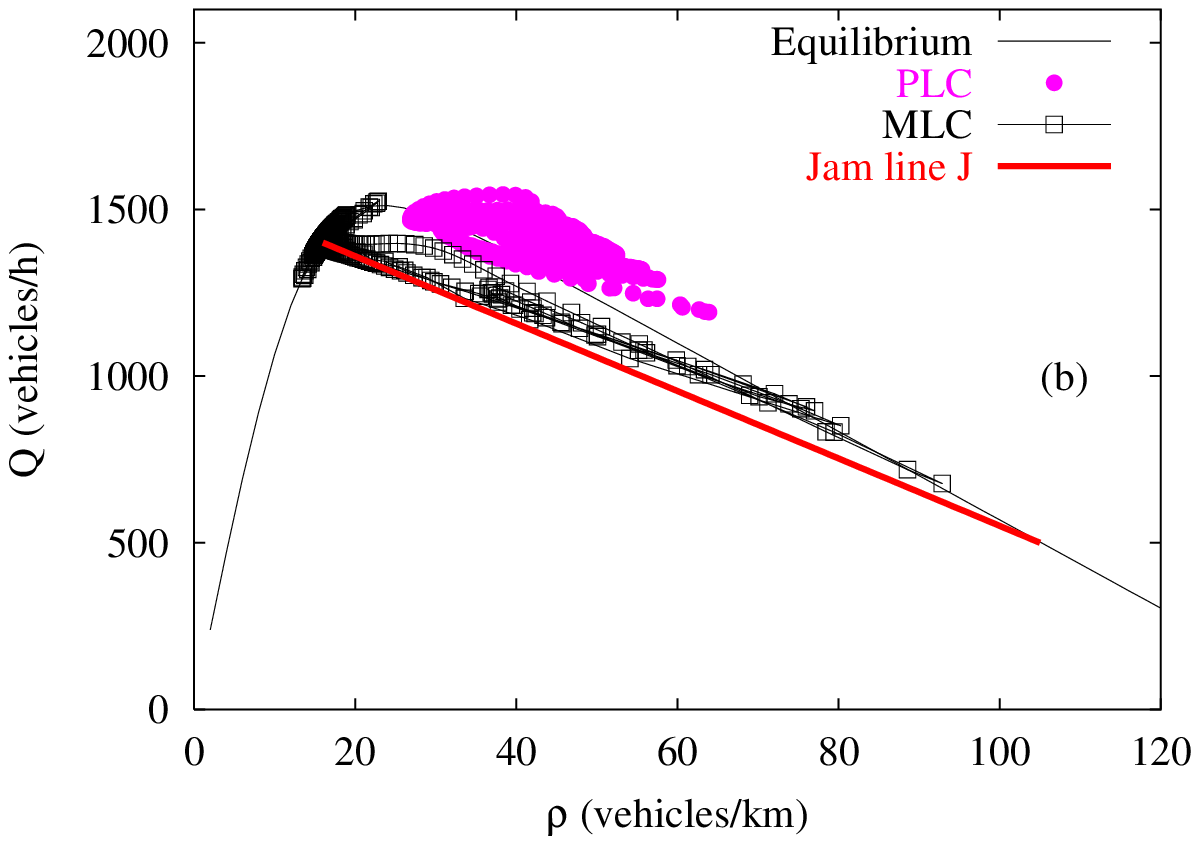}
\end{center}
\caption[]{\label{PLCMLCmic}
{Smoothed spatio-temporal plot of the traffic density
showing the coexistence of a pinned localized cluster (PLC)
with moving localized clusters (MLCs) in an IDM simulation.
The PLC is positioned at a road section with locally {\it increased}
capacity corresponding to a bottleneck strength
$\delta Q=-300$ vehicles/h, which can be identified with the region
between the off- and on-ramps of intersection I2 
in Fig. \protect\ref{PLCMLCemp}(a). It was produced by 
locally {\it decreasing} the IDM parameter
$T$ from 1.9 s to $T'=1.6$ s in a 400 m wide section 
centered at $x=480.8$ km, and increasing it again from $T'$ to $T$ in
a 400 m wide section centered at $x=480.2$ km.
The initial conditions correspond to equilibrium traffic of flow 
$Q_{\rm in}=1390$ vehicles/h, to which a triangular-shaped density peak
(with maximum density 60 vehicles/km at $x=480.5$ km and total width 1 km)
was superposed to initialize the PLC.
As upstram boundary conditions, 
we assumed free equilibrium traffic with
a constant inflow $Q_{\rm in}$ of 1390 vehicles/h.
As downstream boundary conditions for the velocity, 
we used the value for equilibrium free traffic most of
the time. However,
for two five-minute intervals at 14:20 h and 15:40 h, we
reduced the velocity to $v=12$ km/h to initialize the
MLCs.}
%
(b) Flow-density diagram of virtual detectors located at the position
$x=480.2$ km (PLC),  and 2.2 km downstream of it (MLC).
Again, the time intervals, where {density} 
waves passed through the PLC, were
omitted in the {(``virtual'')} data points {belonging to} PLC.
The thin solid curve indicates the equilibrium flow density relation
and 
the thick solid line $J$ characterizes
the outflow from 
fully developed density clusters to free traffic, 
cf. Ref. 
\protect\cite{Kerner-wide}, both for $T=1.9$ s.
}
\end{figure}

Our traffic data indicate that the majority of traffic
breakdowns is triggered by some kind of {\it stationary}
inhomogeneity, so that the
phase diagram is applicable. 
{Such inhomogeneities can be of a very general nature.
They include not only ramps, gradients, lane narrowings or -closings, 
but also incidents in the {\it oppositely} flowing traffic.
In the latter case, the bottleneck is constituted by a
temporary loss of concentration and by braking maneuvers 
of curious drivers {\it at a fixed location}.}
From the more 
than 100 breakdowns on various German and
Dutch freeways investigated by us, there were only four {cases}
(among them the two moving localized clusters
{in} Fig. \ref{PLCMLCemp}), where we could not explain the breakdowns 
by some sort of stationary bottleneck within the 
road sections for which data were available to us.
Possible explanations for the breakdowns in the four remaining cases
are not only spontaneous breakdowns \cite{sync-Letter}, but also
breakdowns triggered by a {\it nonstationary} perturbation,
e.g., moving ``phantom bottlenecks'' caused by two 
trucks overtaking each other \cite{Gazis-Herman}, or
inhomogeneities outside the considered sections. 
Our simulations showed that stationary downstream fronts
are a signature of non-moving
bottlenecks.

\begin{figure}
  \begin{center}
    \includegraphics[width=75mm]{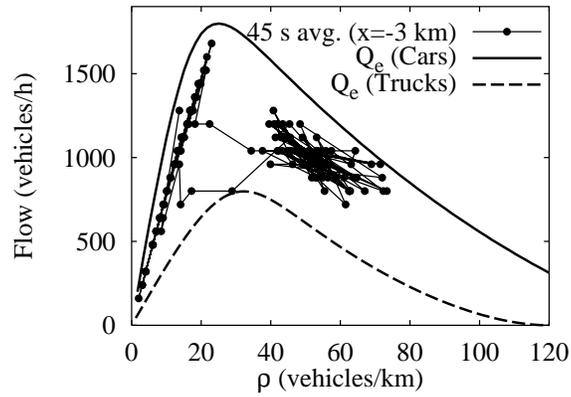}
  \end{center}
\caption{\label{mixfund}
Flow-density diagram of a HCT state of single-lane
heterogeneous traffic consisting of 70\%
``cars'' and 30\% ``trucks''.
Trucks are
characterized by lower IDM parameters $v_0$ and $a$, and a larger
$T$ compared to cars. The solid and dashed curves give
the equilibrium flow-density relations for traffic consisting only of
cars and trucks, respectively.  
For details, see Ref. \protect\cite{TGF99-Treiber}.
}
\end{figure}

Finally, we could qualitatively reproduce the collective dynamics of
{several} rather complex traffic breakdowns by microsimulations
with the IDM, using empirical data for the boundary conditions.
We varied only a {\it single} model parameter, the safe time headway,
to adapt the model to the individual capacities of the different
roads, {\it and} to implement the bottlenecks.
We also performed separate macrosimulations with 
the GKT model and could reproduce the observations
as well. Because both models are effective-single lane models, this
suggests that lane changes are not
relevant {to reproduce} 
the collective dynamics causing the different types of 
congested traffic. 
Furthermore, we assumed identical vehicles and therefore conclude
that also the heterogeneity of real traffic
is not necessary for the basic mechanism of traffic instability.
We expect, however, that other yet unexplained aspects of congested
traffic require
a microscopic treatment of both, multi-lane traffic and heterogeneous 
traffic. These aspects include
the wide scattering of flow-density data
\cite{Kerner-wide,GKT-scatter} {(see Fig.~\ref{mixfund})},
the description of platoon formation \cite{platoon-multilane},
{and the realistic} simulation of speed limits 
\cite{note-phase2-speedlimit},
for which a multi-lane generalization of the
IDM seems to be promising \cite{Applet-engl}.\\[4mm]
{\bf Acknowledgments:}
The authors want to thank for financial support by the BMBF (research
project SANDY, grant No.~13N7092) and by the DFG (grant No.~He 2789). 
We {are also grateful to} the {\it Autobahndirektion S\"udbayern}
and the {\it Hessisches Landesamt f\"ur Stra{\ss}en und Verkehrswesen}
for providing {the} freeway data.



\begin{thebibliography}{10}
\bibitem{note-phase2-micperf}
The simulations were performed on a personal workstation with a 433 Mhz Alpha
  processor.
\end{thebibliography}




\vspace{0mm}

\newcommand{\entry}[2]{\parbox{50mm}{#1} &
                      \parbox{30mm}{#2} }

\begin{table}

\begin{tabular}{l|l}
\entry{Parameter} {Typical\\ value}  \\[3mm] \hline 
\entry{}{}{} \\[-2mm]
\entry{Desired velocity $v_0$}
     {120 km/h}
     \\[0mm]
\entry{Safe time headway $T$}
     {1.6 s}
     \\[0mm]
\entry{Maximum acceleration $a$}
     {0.73 m/s$^2$}
     \\[0mm]
\entry{Desired deceleration $b$}
     {1.67 m/s$^2$}
     \\[0mm]
\entry{Acceleration exponent $\delta$}
     {4}
     \\[0mm]
\entry{Jam distance $s_0$}
     {2 m}
     \\[0mm]
\entry{Jam Distance $s_1$}
     {0 m}
    \\[0mm]
\entry{Vehicle length $l = 1/\rho_{\rm max}$}
     {5 m}
\end{tabular}
\vspace*{5mm}
\caption{\label{tab:param} Model parameters of the IDM model
used throughout this paper. {Changes of the freeway capacity were
described by a variation of the safe time headway $T$.}
}
\end{table}


\newpage

\newpage
\newpage

\newpage




\newpage


\newpage


\newpage



\newpage


\newpage


\newpage


\newpage



\newpage


\newpage
\newpage


\newpage


\end{document}